\documentclass{aa}  

\usepackage{graphicx}
\usepackage{comment}
\usepackage{pifont}

\usepackage{txfonts}

\usepackage{natbib}
\bibpunct{(}{)}{;}{a}{}{,} 

\usepackage{xpatch}
\usepackage{hyperref}
\usepackage{placeins}
\hypersetup{
	colorlinks=true,
	breaklinks=true,
	citecolor=blue,
	allcolors=blue,
	frenchlinks=true
}

\makeatletter
\xpatchcmd\NAT@citex
{
	\@citea\NAT@hyper@{
		\NAT@nmfmt{\NAT@nm}
		\hyper@natlinkbreak{\NAT@aysep\NAT@spacechar}{\@citeb\@extra@b@citeb}
		\NAT@date
	}
}
{
	\@citea
	\NAT@nmfmt{\NAT@nm}
	\NAT@aysep\NAT@spacechar
	\NAT@hyper@{\NAT@date}
}
{}{}
\xpatchcmd\NAT@citex
{
	\@citea\NAT@hyper@{
		\NAT@nmfmt{\NAT@nm}
		\hyper@natlinkbreak{\NAT@spacechar\NAT@@open\if*#1*\else#1\NAT@spacechar\fi}
		{\@citeb\@extra@b@citeb}
		\NAT@date
	}
}
{
	\@citea
	\NAT@nmfmt{\NAT@nm}
	\NAT@spacechar\NAT@@open\if*#1*\else#1\NAT@spacechar\fi
	\NAT@hyper@{\NAT@date}
}
{}{}

\defcitealias{Schneider20}{S20}
\newcommand{\labsim}[1]{\fontfamily{qcr}\selectfont \textbf{#1}}
\begin{document}

   \title{Modeling Multiphase Galactic Outflows: A Multifluid Moving Mesh Approach}

   \author{F. Bollati
          \and
          R. Weinberger
          }

   \institute{Leibniz Institute for Astrophysics Potsdam (AIP), An der Sternwarte 16, 14482 Potsdam, Germany\\
              \email{fbollati@aip.de}
             }

   \date{}

  \abstract{  
  Outflows are a key part of the galactic gas cycle and crucial in shaping the star formation activity in their host galaxy. Yet, in simulations of galaxy evolution, modeling these outflows in their multi-phase nature and over the relevant  timescales is an unsolved problem.
  We present a subgrid model for simulating multiphase galactic outflows in efficient, comparatively low-resolution simulations, designed for application in future cosmological simulations. The cold phase ($T=10^4$~K) is treated as pressureless, and its interaction with the hot phase is captured through source terms representing drag and mixing. These terms are obtained using analytic drag and mixing terms for single clouds and convolving them with a cloud mass distribution consistent with high-resolution simulations. Applied to a setup resembling the starburst galaxy M82, the model reproduces the velocity, density, and mass outflow rates of high-resolution simulations that resolve individual cold clouds. Cold outflows emerge naturally from interactions between the hot wind and cold interstellar clouds, with drag and mixing both contributing to the acceleration. Varying the mixing strength strongly affects outflow properties: stronger mixing enhances mass transfer from hot to cold gas, reduces the hot phase velocity, and accelerates the cold phase, while also influencing the origin and composition of the cold outflow. Weak mixing produces cold gas mostly from preexisting interstellar clouds, whereas stronger mixing leads to substantial cold gas formation from the hot phase. This framework enables efficient simulations of multiphase galactic outflows while retaining key multi-component features of the outflow dynamics.
  }


   \maketitle
%

\section{Introduction}

Galactic winds play a fundamental role in galaxy evolution as they expel gas from the host galaxy, heat and enrich the circum-galatic medium with metals and regulate star formation in the host galaxy.  
However, despite their perceived importance, the many details related to galactic winds are still unclear: which physical processes drive galactic outflows? How much mass and energy is actually removed from galaxies? Which phase dominates the outflow? How universal are these properties as a function of galaxy mass, morphology and redshift? Answering these questions is key to a more complete understanding of galaxy formation (see \citealt{Thompson24} for a recent review).  

From an observational perspective, galactic outflows are detected both in the local universe 
\citep[e.g.,][]{Heckman90} and at high redshift 
\citep[e.g.,][]{Shapley03, Genzel14, Fosterschreiber19, Guo23, Belli24}, 
establishing them as a common feature of star-forming galaxies. 
A key finding is that these winds are multiphase in nature, spanning an enormous range of gas temperatures from $\sim 10$ K to $\gtrsim 10^7$ K, and traced through both emission and absorption features.  
The cold molecular phase ($T \sim 10$ K) is revealed by tracers such as CO 
\citep[e.g.,][]{Bolatto13, Leroy15, Veilleux20, Fluetsch21} 
and PAH emission \citep[e.g.,][]{Fisher25}.  
The cool neutral phase ($T \sim 10^2$ K) is detected via H\,\textsc{i} 
\citep[e.g.,][]{Martini18, Watts24} 
and low-ionization lines such as Na\,D, O\,\textsc{i}, and Mg\,\textsc{i} 
\citep{Rupke05a, Davies24}.  
A warm ionized phase ($T \sim 10^{4-5}$ K) is traced by H$\alpha$ 
\citep[e.g.,][]{McKeith95, Westmoquette09} 
and metal lines including Mg\,\textsc{ii}, O\,\textsc{ii}, O\,\textsc{iii},and Si\,\textsc{iv} 
\citep[e.g.,][]{Rubin11, Chisholm17, Concas19, Ha25}.  
The warm–hot phase ($T \sim 10^{5-6}$ K) is detected in higher ionization tracers such as O\,\textsc{vi} 
\citep[e.g.,][]{ Ashley20, Ha25},  
while the hottest gas ($T \gtrsim 10^{6}$ K) is observed in diffuse X-ray emission 
\citep[e.g.,][]{Strickland09, Lopez20}.  
Nearby archetypal starbursts such as M82 and NGC 253 show evidence for all of these phases coexisting in the same outflow 
\citep[e.g.,][]{Bolatto13, Lopez20},  
while at higher redshift typically only a subset of phases are detected, often limited to ionized or neutral gas 
\citep[e.g.,][]{Davies24, Watts24, Belli24}.  
Despite these advances, significant uncertainties remain: the relative mass-loading factors of different phases, their energy and momentum budgets, and the physical mechanisms that generate and maintain multiphase structure are still poorly constrained observationally. Addressing these open questions increasingly relies on numerical simulations, which can disentangle the role of different processes in shaping the multiphase outflow.

From a theoretical perspective, simulating multiphase outflows resolving even the $T \sim 10^4$ K phase has been proven challenging. Cloud crushing simulations  \citep[e.g.,][]{Klein94, Xu95} provide a controlled setup to study at high-resolution the interaction between a single cold ($T \sim 10^4$~K) cloud and the hot wind in which it is embedded. The general picture emerging from cloud–wind interaction simulations is that the relative motion between clouds and the hot flow excites hydrodynamical instabilities—primarily Kelvin–Helmholtz and Rayleigh–Taylor instabilities—at the phase interface. These instabilities generate turbulence and create a mixing layer between the two phases, characterized by intermediate temperatures \citep{Gronke18, Fielding20a}. Within this layer, two competing processes operate: turbulent mixing, which strips material from and can ultimately destroy the clouds, and radiative cooling, which can condense hot material onto the cloud surface and promote cloud mass growth \citep[e.g.,][]{Gronke18, Gronke20, Fielding20a, Tan21, Kanjilal21, Abruzzo22}. Both processes are accompanied by associated momentum and energy exchange, while in parallel the clouds are accelerated by ram pressure drag. The outcome of cloud survival or destruction is therefore set by the relative timescales of these competing effects \citep{Fielding22}.  

A large body of work has explored how additional physical processes alter the cloud–wind interaction, including conduction \citep[e.g.,][]{Bruggen16}, magnetic fields \citep[e.g.,][]{McCourt15, Cottle20, Sparre20}, cosmic rays \citep[e.g.,][]{Wiener19, Bruggen16}, viscosity \citep[e.g.,][]{Marin-Gilabert25}, the pressure gradients of expanding winds \citep[e.g.,][]{Dutta25}, and turbulence in the hot medium \citep[e.g.,][]{Das2024, Marin-Gilabert25}. More recent studies have begun to examine populations of clouds embedded in hot flows or turbulent environments. These works suggest that the collective interaction is not a simple superposition of individual cloud–wind encounters, but instead gives rise to more complex emergent behaviors \citep[e.g.,][]{Villares24, Tan24, Seidl25}.

On larger scales, ``tall-box'' simulations have been employed to bridge galactic and cloud-scale studies \citep[e.g.,][]{Girichidis16, Martizzi16, Simpson2016, Fielding18, Kim18, Kim20, Girichidis21, Ostriker22, Armillotta24}. These simulations follow kiloparsec-scale patches of the ISM with sufficient resolution to capture the dynamics of small cold clouds embedded in a hot wind. They self-consistently model the injection of energy and momentum from young stellar populations, including stellar winds, radiation pressure, and core-collapse supernovae, and track the interaction of this feedback with the multiphase ISM. Because of computational constraints, typical resolutions are of order $\Delta m \sim 10^{-3}-10 \,\mathrm{M}_\odot$ or $\Delta x \sim$ pc, limiting the domain size to a few kpc. 

On even larger scales, individual galaxy simulations complement tall-box studies by allowing the wind to expand into its full biconical structure and by capturing more complex global phenomena such as galactic fountains. These simulations typically achieve coarser resolution, with $\Delta m \sim 10$--$10^3 \, \mathrm{M}_\odot$ and $\Delta x \sim 1$--$100$ pc \citep[e.g.,][]{Tanner16, Emerick18, Schneider18, Smith18, Vijayan18, Hu19, Schneider20, Steinwandel2024, Thomas25}. Despite these limitations, they are able to self-consistently model both the launching of galactic winds and their subsequent multiphase evolution, resolving the development of complex cold cloud structures and their interaction with the hot outflow.

Cosmological simulations, owing to the vast spatial and temporal scales involved, necessarily lack the resolution (typically $\Delta m \sim 5 \times 10^3$--$10^5 \, \mathrm{M}_\odot$, $\Delta x \sim \textrm{few}$--$100$ pc for zoom-in runs) required to capture the detailed small-scale processes in the ISM that drive galactic winds, as well as the internal structure of the winds themselves \citep[see][]{Naab17}. As a result, feedback processes from stars (and AGN) must be modeled through subgrid prescriptions that inject energy and momentum into the surrounding medium. The free parameters of these models are usually calibrated against key observables 
\citep[e.g.,][]{Vogelsberger13, Crain15, Pillepich18}, and/or higher-resolution simulations \citep[e.g.,][]{Dave19}.  
However, even state-of-the-art cosmological simulations generally fail to resolve the multiphase structure of galactic winds. Only in recent high-resolution runs it become possible to marginally resolve the formation of the most massive clouds entrained in outflows \citep{Nelson19, Mitchell20, Pandya21}, though these still fall short of reproducing the full cold cloud population seen in observations.  

To overcome these limitations, new techniques have been developed to explicitly track the behavior of unresolved cold clouds within large-scale simulations. One approach is to generalize the ``wind particle" approach \citep{Springel2003} to allow for multiple species \citep{Dave2016},  evolution \citep{Huang2020, Huangetal2022} into a de-facto particle based pressureless fluid component including mass exchange and drag terms \citep{Smith24}.
An alternative, grid-based approach is presented in \citet{Butsky24}, where they implemented the analytic cloud–wind interaction prescriptions of \citet{Fielding22} into the \textsc{Enzo} code to model unresolved cloud evolution in the circumgalactic medium. Following this philosophy, we present a new model that builds on the multi-fluid implementation of \textsc{Arepo} by \citet{Weinberger23}, extending it with a subgrid treatment of the interaction between the hot wind and unresolved cold clouds and study its behavior on individual galaxy simulations of a starburst-driven galactic outflow.

The paper is structured as follows. In Section~\ref{Sec: Model}, we introduce our model for the unresolved cloud population. We describe the simulation setup and the suite of simulations performed in this work in Section~\ref{sec: M82}. Sections~\ref{sec: results} covers the main results which are discussed in Section~\ref{sec: discussion}. We summarize our findings in Section~\ref{sec: conclusion}.

\section{Model} \label{Sec: Model}

In this section, we introduce a new model for unresolved, cold clouds based on the multi-fluid framework introduced by \citet{Weinberger23}, which enables us to track their interaction with the surrounding, hotter medium. In the multi-fluid discretization of \textsc{Arepo}, each resolution element contains the conserved quantities — mass $M$, momentum $\mathbf{P}$, and energy $E$ — of multiple gas phases; in our case, two: a cold phase at $T_\textrm{cold} = 10^4$ K and a hot phase with $T_\textrm{hot} \geq 10^4$~K.
Each gas phase evolves according to fluxes exchanged with neighboring cells. Notably, for the simulations presented in this work we simplify the multi-fluid treatment by assuming the cold phase to be pressureless, meaning its pressure is formally set to zero in the multi-fluid hydrodynamic equations. In this limit, our framework reduces to the model of \citet{Butsky24} or grid-based implementations of dust \citep[e.g.][]{Huang2022}. The approximation is adequate as long as the volume filling fraction of the cold gas is small, and has the practical advantage of being numerically simpler and trivially generalizable to multi-species flows \citep{Benitez-Llambay2019}. The different phases within a cell can interact via source terms representing mass exchange and drag forces. These source terms are integrated in an operator-split fashion using the semi-implicit integrator suggested by \citet{Bader1983}. A forthcoming paper (Weinberger et al., in prep.) will cover the detailed implementation in the \textsc{Arepo} code.

\begin{figure}
    \centering
    \includegraphics[width=\linewidth]{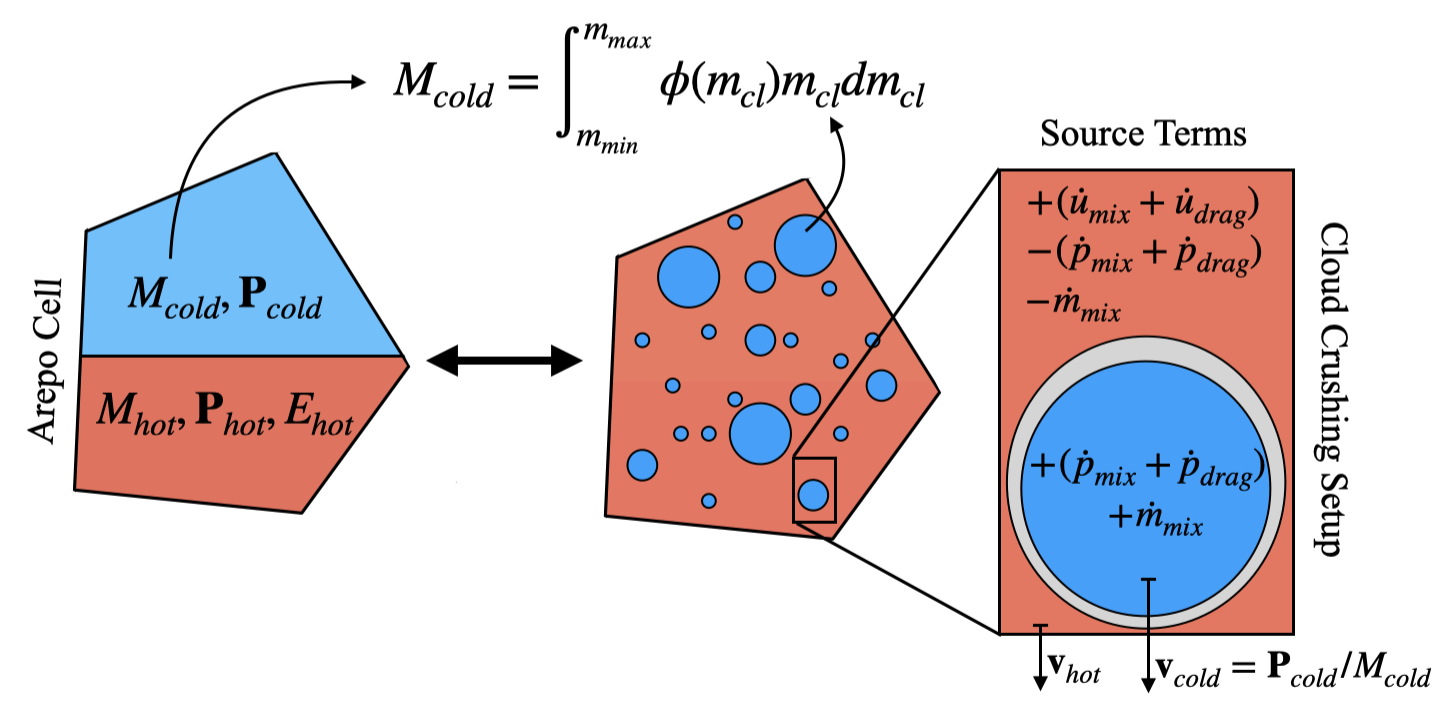}
    \caption{On the left, a single \textsc{Arepo} cell contains conserved quantities for both a hot phase and a cold, pressureless phase. The interaction between the two phases within the cell is modeled by assuming that the cold phase consists of a population of clouds with a mass function $\phi$, embedded in the hot medium. Each cloud interacts individually with the surrounding hot environment, following the dynamics predicted by cloud-crushing simulations. In particular, mixing and drag processes result in the exchange of mass, momentum, and energy between the two phases. The total source terms for the two phases are obtained by summing the contributions from all individual clouds. All the source terms appearing in the cloud crushing box on the right are the single cloud source terms defined in Eqs. (\ref{Eq: mdot mix}, \ref{Eq: pdot mix}-\ref{Eq: dotu drag}).}
    \label{Fig: scheme}
\end{figure}

In this work we focus on the inclusion of sub-grid interaction terms between different gas phases within the same resolution element, and its consequences for galactic outflows. These interaction terms are based on the assumption that the cold phase consists of small clouds embedded in a hot medium, and that each cloud interacts with the hot phase as predicted by high-resolution cloud-crushing simulations (e.g., \citealt{Gronke18, Gronke20, Fielding20a, Tan21,Kanjilal21, Abruzzo22}). A schematic representation of our model is shown in Figure~\ref{Fig: scheme}. Cloud-crushing simulations typically model a single, spherical, homogeneous cold cloud impacted by a hot wind, tracking the system evolution over a few cloud-crushing times—that is, a few multiples of the characteristic timescale for the cloud to be disrupted and mixed by the surrounding flow. These simulations show that the interaction between the cloud and the hot wind is governed primarily by two processes: turbulent mixing at an intermediate-temperature layer that forms between the two phases, and a drag (ram-pressure) force exerted by the wind on the cloud. Building on these results, \citet{Fielding22} derived simple analytical expressions for the exchange of mass, momentum, and energy between a cold cloud and the impinging hot wind. Below, we briefly summarize these prescriptions and explain how we incorporate them to describe the interaction between gas phases within an individual multi-fluid \textsc{Arepo} cell.

Let us consider a cold cloud of mass $m_\textrm{cl}$ and radius $r_\textrm{cl}$, moving with velocity $v_\textrm{rel}$ relative to a hot ambient phase. At the interface between the two phases, hydrodynamical instabilities develop, giving rise to turbulence within the intermediate mixing layer. This turbulence entrains material from both phases, progressively shredding the cold cloud and causing it to lose mass at a rate
\begin{equation}
\dot{m}_\textrm{loss} = 3f_\textrm{mix} \frac{m_\textrm{cl} f_\textrm{turb}v_\textrm{rel}}{\chi^{1/2} r_\textrm{cl}},
\label{Eq: mdot mix loss}
\end{equation}
where $f_\textrm{turb}v_\textrm{rel}$ denotes the turbulent velocity within the layer, $\chi = \rho_\textrm{cl}/\rho_\textrm{hot}$ is the density contrast between the cloud and the hot medium, and $f_\textrm{mix}$ is a numerical factor representing the effective cloud area over which mixing occurs \citep{Abruzzo22}. This area increases beyond the initial spherical surface of the cloud as it becomes elongated during the interaction. As pointed out by \cite{Gronke18}, a cold cloud can also gain mass, since the turbulent mixing layer contains material at an intermediate temperature between the cold and hot phases, where radiative cooling is particularly efficient. As a result, some of the gas in this layer can cool down to the cold-phase temperature $T_\textrm{cold}$ and accrete onto the cloud at a rate
\begin{equation}
\dot{m}_\textrm{growth} = \dot{m}_\textrm{loss} \cdot \xi^\alpha,
\label{Eq: mdot mix grwoth}
\end{equation}
where
\begin{equation}
\xi = \frac{r_\textrm{cl}}{t_\textrm{cool} f_\textrm{turb} v_\textrm{rel}},
\label{Eq: xi}
\end{equation}
and the exponent $\alpha$ depends on the value of $\xi$: specifically, $\alpha = 1/4$ if $\xi \geq 1$ and $\alpha = 1/2$ otherwise \cite[see e.g.][]{Fielding20a}. The cooling timescale $t_\textrm{cool}$ refers to the gas in the mixing layer, which is assumed to have a characteristic temperature $T_\textrm{layer} = \sqrt{T_\textrm{hot} T_\textrm{cloud}}$ \citep[e.g.][]{Gronke18, Fielding22}. Combining Eqs. (\ref{Eq: mdot mix loss}) and (\ref{Eq: mdot mix grwoth}), the net rate of change in the cloud mass is given by
\begin{equation}
\dot{m}_\textrm{cl, mix} = \dot{m}_\textrm{ growth} - \dot{m}_\textrm{loss} = \dot{m}_\textrm{loss} \bigl( \xi^\alpha - 1 \bigr).
\label{Eq: mdot mix}
\end{equation}
From Eqs.(\ref{Eq: xi}) and (\ref{Eq: mdot mix}), it follows that there exists a critical cloud radius for which $\xi = 1$, namely
\begin{equation}
r_\textrm{crit} = t_\textrm{cool} f_\textrm{turb} v_\textrm{rel},
\label{Eq: r_crit}
\end{equation}
such that clouds with $r_\textrm{cl} > r_\textrm{crit}$ experience net growth (with larger clouds growing more rapidly), whereas those with $r_\textrm{cl} < r_\textrm{crit}$ undergo net mass loss. The mass exchange described above is inherently accompanied by momentum and energy transfer between the two phases, since the exchanged mass carries its own momentum and energy. As a result, the evolution of the cloud momentum and of the internal energy of the hot phase can be expressed as
\begin{align}
&\dot{\mathbf{p}}_\textrm{cl,mix} = \dot{m}_\textrm{growth} \cdot \mathbf{v}_\textrm{hot} - \dot{m}_\textrm{loss} \cdot \mathbf{v}_\textrm{cold}, \label{Eq: pdot mix} \\
&\dot{u}_\textrm{cl,mix} = \frac{\dot{m}_\textrm{loss}}{M_\textrm{hot}} \left[ \left( u_\textrm{cold} - u_\textrm{hot} \right) + \frac{1}{2}v_\textrm{rel}^2 \right],
\label{Eq: udot mix}
\end{align}
where $\mathbf{v} $ and $u$ denote the velocity and specific internal energy of the phase indicated by the subscript.
We stress that since the cold phase is treated as pressureless in this work, the thermalization of relative kinetic energy associated with mass exchange  is entirely attributed to the hot phase.

In addition to mass exchange, the cloud is also subject to a ram pressure force exerted by the hot wind, given by
\begin{equation}
\dot{\mathbf{p}}_\textrm{cl,drag} = \frac{\pi}{4} r_\textrm{cl}^2 \,\rho_\textrm{hot} v_\textrm{rel} \cdot \mathbf{v}_\textrm{rel} \equiv K_\textrm{cl} \cdot \mathbf{v}_\textrm{rel},
\label{Eq: dotp drag}
\end{equation}
where $K_\textrm{cl}$ represents the drag coefficient. This momentum transfer is associated with a corresponding increase in the internal energy of the hot phase, given by
\begin{equation}
\dot{u}_\textrm{cl,drag} = \frac{K_\textrm{cl}}{M_\textrm{hot}} v_\textrm{rel}^2.
\label{Eq: dotu drag}
\end{equation}

Equations~(\ref{Eq: mdot mix}, \ref{Eq: pdot mix}, \ref{Eq: udot mix}, \ref{Eq: dotp drag}, \ref{Eq: dotu drag}) define the source terms governing the mutual interaction between the cold cloud and the hot wind. In the following, we incorporate these terms as subgrid source terms describing the exchange of mass, momentum, and energy between the hot and cold phases within an \textsc{Arepo} cell.
To enable this treatment, we adopt a subgrid model in which the cold phase is composed of a population of small cold clouds that interact with the surrounding hot medium (i.e., the hot component of the cell), consistent with the behavior observed in cloud-crushing simulations. In practice, this approach requires an assumption about how the total cold-phase mass in the cell is distributed among individual cold clouds. Below, we present two alternative prescriptions for this decomposition.

\subsection{Cloud model 1: delta distribution} \label{sec: delta}

The simplest modeling approach assumes that the cold phase in each simulation cell consists of clouds of uniform size, denoted by the free parameter $r_\textrm{cl}$. The mass of each cloud is a local quantity (i.e., varying from cell to cell), and is given by
\begin{equation}
m_\textrm{cl} = \frac{4}{3}\pi r_\textrm{cl}^3 \rho_\textrm{cl},
\label{Eq: mcl}
\end{equation}
where the cloud density $\rho_\textrm{cl}$ is determined by the pressure equation:
\begin{equation}
\rho_\textrm{cl} = \varepsilon_P P_\textrm{hot} \frac{\mu m_\textrm{p}}{k_\textrm{B} T_\textrm{cold}}.
\label{Eq: rhocl}
\end{equation}
Here, $P_\textrm{hot}$ is the pressure of the hot phase, $m_\textrm{p}$ is the proton mass, $\mu$ the mean molecular weight, $k_\textrm{B}$ the Boltzmann constant, and $\varepsilon_P$ a coefficient representing the ratio of cold- to hot-phase pressure, introduced to account for deviations from pressure equilibrium. Under this assumption, the cloud mass function is, within a single cell, a delta distribution:
\begin{equation}
\phi(m) \equiv \frac{dN}{dm} = N_\textrm{cl} \, \delta(m - m_\textrm{cl}),
\end{equation}
where the number of cold clouds in a cell is given by
\begin{equation}
N_\textrm{cl} = \frac{M_\textrm{cold}}{m_\textrm{cl}}.
\label{Eq: Ncl}
\end{equation}
All clouds within a cell are assumed to share the same velocity, determined by the macroscopic (of the cell) properties of the cold phase: $\mathbf{v}_\textrm{cold} = \mathbf{P}_\textrm{cold} / M_\textrm{cold}$.
With this formulation, the source terms for the hot phase in a given cell are:
\begin{align}
&\dot{M}_\textrm{hot}^\textrm{d} = -N_\textrm{cl} \cdot \dot{m}_\textrm{cl, mix}, \label{Eq: umd Mdot}\\
&\dot{\mathbf{P}}_\textrm{hot}^\textrm{d} = -N_\textrm{cl} \cdot \left( \dot{\mathbf{p}}_\textrm{cl,mix} + \dot{\mathbf{p}}_\textrm{cl,drag} \right), \label{Eq: umd Pdot}\\   
&\dot{u}_\textrm{hot}^\textrm{d} =  N_\textrm{cl} \cdot \left( \dot{u}_\textrm{cl,mix} + \dot{u}_\textrm{cl, drag} \right) + \dot{u}_\textrm{cool}. 
\label{Eq: umd udot}
\end{align}
Here superscript 'd' stands for delta distribution. 
The rates of change for the cold-phase mass and momentum are simply the negatives of those in Eqs.~(\ref{Eq: umd Mdot},~\ref{Eq: umd Pdot}).  In this model, the source terms for the phases in a given cell are just those of a single cloud times the number of clouds per cell. Equation~(\ref{Eq: umd udot}) also includes the term $\dot{u}_\textrm{cool}$, which represents the contribution from radiative cooling processes occurring within the hot phase and is unrelated to interactions with the cold phase.

We can now easily derive the scaling relations of the cell source terms with cloud mass $m_\textrm{cl}$. In the regime where $\xi \ll 1$, cloud mass exchange is dominated by mass loss. From Eq.~(\ref{Eq: mdot mix loss}) and using the scaling $r_\textrm{cl} \propto m_\textrm{cl}^{1/3}$, we obtain $\dot{m}_\textrm{cl,mix} \propto m_\textrm{cl}^{2/3}$. In the opposite regime, $\xi \gg 1$, mixing is dominated by mass growth, and from Eq.~(\ref{Eq: mdot mix grwoth}), $\dot{m}_\textrm{cl,mix} \propto m_\textrm{cl}^{3/4}$. 
Combining these scalings with Eq.~(\ref{Eq: Ncl}) and ~(\ref{Eq: umd Mdot}), we find that the total mass exchange rate with the hot phase scales as $\dot{M}_\textrm{hot} \propto m_\textrm{cl}^{-1/3}$ for $\xi \ll 1$, and $\dot{M}_\textrm{hot} \propto m_\textrm{cl}^{-1/4}$ for $\xi \gg 1$. Similarly, from Eq.~(\ref{Eq: dotp drag}) and ~(\ref{Eq: Ncl}), the total drag force scales as $N_\textrm{cl} \dot{\mathbf{p}}_\textrm{drag} \propto m_\textrm{cl}^{-1/3}$. In other words, the overall mass exchange, and the associated momentum and energy transfer, as well as the drag-induced momentum exchange, and its corresponding energy exchange, all decrease when the cold phase is composed of a few massive clouds, and increase when it consists of many small clouds. This behavior will be important when comparing this single-cloud-mass model to the one presented in the next subsection.
Note that throughout this paper, we use this delta distribution model only as a comparison in the appendix. The fiducial cloud model is the piecewise power law model.

\subsection{Cloud model 2: piecewise power law mass distribution} \label{sec: ppl}

The second model relaxes the assumption of single-mass clouds within a cell and instead adopts a piecewise power-law mass function for the clouds. This approach is motivated by high-resolution simulations of multiphase outflows \citep{Tan24, Warren25}, which are capable of resolving cold clouds and directly measuring their mass distribution. We assume that the cloud mass function within a resolution element is given by
\begin{equation}
    \phi(m) = \mathcal{N}
    \begin{cases} 
     m_\textrm{crit}^{-2}(m/m_\textrm{crit})^{-\gamma} & \text{if} \quad m < m_\textrm{crit},\\
     m^{-2} & \text{if} \quad m \geq m_\textrm{crit},
    \end{cases}
    \label{Eq: clouds distribution}
\end{equation}
where $m_\textrm{crit}$ is the mass of a cloud with critical size $r_\textrm{crit}$, and the function is defined over the range $m \in (m_\textrm{min}, m_\textrm{max})$. The parameters $\gamma$, $m_\textrm{min}$, and $m_\textrm{max}$ are free parameters of the model. The normalization constant $\mathcal{N}$, set by $\int_{m_\textrm{min}}^{m_\textrm{max}} m\, \phi(m)\, dm = M_\textrm{cold}$, reads
\begin{equation}
\mathcal{N} =  M_\textrm{cold}\Biggl[ \frac{1-(m_\textrm{min}/m_\textrm{crit})^{2-\gamma}}{2-\gamma} + \ln\Biggl(\frac{m_\textrm{max}}{m_\textrm{crit}}\Biggr) \Biggr]^{-1}.
\label{Eq: Normalization phi}
\end{equation}
We remark that the two slopes of the mass distribution, $-\gamma$ and $-2$, are assumed to be constant. However, Eqs.~(\ref{Eq: mdot mix loss}) and ~(\ref{Eq: mdot mix grwoth}) indicate that the mass exchange rate of a single cloud depends on the cloud mass: larger clouds tend to gain mass more rapidly, while smaller clouds lose mass more slowly. This differential interaction across the cloud population would, in principle, cause the distribution to evolve away from its initial shape over time.
Nevertheless, high-resolution simulations of multiphase outflows, such as those by \citet{Tan24} and \cite{Warren25}, show that the overall shape of the cloud mass distribution remains stable throughout the simulation. This suggests that additional physical processes are at play, beyond individual cloud–hot phase interactions, such as cloud mergers and fragmentation. These processes, while not directly contributing to the mass exchange with the hot phase, play a key role in preserving the slopes of the mass distribution.
In our model, such effects are not explicitly included. Instead, we effectively account for them by assuming that the cloud mass function retains a fixed shape over time.
Similarly, through processes such as growth and mass loss, clouds in the population could, in principle, move beyond the defined mass boundaries, $m_\textrm{min}$ and $m_\textrm{max}$. However, the discussion above implies that fragmentation and mergers tend to keep cloud masses within this prescribed range, although this may be somewhat debatable. Nevertheless, as detailed in the Appendix~\ref{app: fractional}, for slope parameters $\gamma \lesssim 1.5$, the specific choice of $m_\textrm{min}$ and $m_\textrm{max}$ has a negligible impact on the computation of source terms. 

Once the cloud mass distribution is defined (Eq. \ref{Eq: clouds distribution}) and the source terms for individual clouds are known (Eqs. \ref{Eq: mdot mix} and \ref{Eq: dotp drag}), the total source terms for the cold phase, represented by the entire cloud population, can be obtained by integrating the single-cloud source terms over the cloud mass function. For instance, the total mass-loss source term for the cold phase is given by
\begin{align}
  \dot{M}_\textrm{loss} & = \int _{m_\textrm{min}}^{m_\textrm{max}}\phi(m) \cdot  \dot{m}_\textrm{loss} \, dm =\\
  & = 3 f_\textrm{mix} \frac{\mathcal{N} f_\textrm{turb} v_\textrm{rel}}{\chi ^{1/2}  r_\textrm{loss}},
 \label{Eq: M cold pop loss} 
\end{align}
where the effective cloud radius $r_\textrm{loss}$ is defined as 
\begin{equation}
    r_\textrm{loss}^{-1} =  \frac{9}{r_\textrm{crit}}\frac{2-\gamma}{5-3\gamma} -\frac{3}{r_\textrm{min}(5-3\gamma)}\Biggl(\frac{m_\textrm{crit}}{m_\textrm{min}}\Biggr)^{\gamma-2}- \frac{3}{r_\textrm{max}}.
    \label{Eq: rloss}
\end{equation}
Similarly, the growth rate of the total cold phase in the cell is found to be 
 \begin{equation}
  \dot{M}_\textrm{growth} = 3f_\textrm{mix} \frac{\mathcal{N}f_\textrm{turb}v_\textrm{rel}}{\chi ^{1/2}r_\textrm{growth}}, 
   \label{Eq: M cold pop growth} 
 \end{equation}
where 
\begin{equation}
r_\textrm{growth}^{-1} =  \frac{6}{11-6\gamma}\Biggl(\frac{1}{r_\textrm{crit}} - \frac{\xi_\textrm{min}^{1/2}}{r_\textrm{min}}\Biggl(\frac{m_\textrm{crit}}{m_\textrm{min}}\Biggr)^{\gamma-2}\Biggr) -4\Biggl( \frac{\xi_\textrm{max}^{1/4}}{r_\textrm{max}} - \frac{1}{r_\textrm{crit}}\Biggr).
\label{Eq: rgrowth}
\end{equation}
Then, analogous to Eqs.~(\ref{Eq: pdot mix}, \ref{Eq: udot mix}) for a single cloud, the total rate of change of momentum in the cold phase and the rate of change of internal energy in the hot phase due to mixing are given by
\begin{align}
&\dot{\mathbf{P}}_\textrm{mix} = \dot{M}_\textrm{growth} \cdot \mathbf{v}_\textrm{hot} - \dot{M}_\textrm{loss} \cdot \mathbf{v}_\textrm{cold}, \label{Eq: pdot pop mix} \\
&\dot{u}_\textrm{mix} = \frac{\dot{M}_\textrm{loss}}{M_\textrm{hot}} \left[ \left( u_\textrm{cold} - u_\textrm{hot} \right) + \frac{1}{2}v_\textrm{rel}^2 \right].
\label{Eq: udot pop mix}
\end{align}
Applying the same procedure, the total drag force acting on the cold phase is given by
\begin{equation}
    \dot{\mathbf{P}}_\textrm{drag} = \frac{\mathcal{N}}{r_\textrm{drag}} \frac{\pi}{4} \, v_\textrm{rel} \cdot \mathbf{v}_\textrm{rel} \equiv K_d \cdot \mathbf{v}_\textrm{rel},
\label{Eq: K pop}
\end{equation}
where 
\begin{equation}
    r^{-1}_\textrm{drag} = \rho_\textrm{hot} \Biggl[ \frac{3}{5-3\gamma}\Biggl(\frac{r_\textrm{crit}^2}{m_\textrm{crit}} - \frac{r_\textrm{min}^2}{m_\textrm{min}}\Biggl(\frac{m_\textrm{crit}}{m_\textrm{min}}\Biggr)^{\gamma-2} \Biggr) - 3\Biggl(\frac{r_\textrm{max}^2}{m_\textrm{max}}-\frac{r_\textrm{crit}^2}{m_\textrm{crit}} \Biggr)\Biggr].
    \label{Eq: rdrag}
\end{equation}
Analogously to Eq.~(\ref{Eq: dotu drag}) for a single cloud, the total change in internal energy of the hot phase due to the drag force acting on the cloud population is given by
\begin{equation}
\dot{u}_\textrm{drag} = \frac{K_\textrm{d}}{M_\textrm{hot}} v_\textrm{rel}^2.
\label{Eq: dotu pop drag}
\end{equation}
We can now express the total source terms for the hot phase as
\begin{align}
&\dot{M}_\textrm{hot}^\textrm{ppl} = \dot{M}_\textrm{loss} - \dot{M}_\textrm{growth}, \label{Eq: ppl Mdot}\\
&\dot{\mathbf{P}}_\textrm{hot}^\textrm{ppl} =  -\dot{\mathbf{P}}_\textrm{mix} - \dot{\mathbf{P}}_\textrm{drag} , \label{Eq: ppl Pdot}\\   
&\dot{u}_\textrm{hot}^\textrm{ppl} =  \left( \dot{u}_\textrm{mix} + \dot{u}_\textrm{drag} \right) + \dot{u}_\textrm{cool}. 
\label{Eq: ppl udot}
\end{align}
Here, the superscript ``ppl'' denotes the piecewise power-law distribution. As in the case of the delta distribution (see Sec.~\ref{sec: delta}), the rates of change for the cold-phase mass and momentum are simply the negatives of those given in Eqs.~(\ref{Eq: ppl Mdot},~\ref{Eq: ppl Pdot}).
In practice, when employing this model, we encounter the issue that $m_\textrm{crit}$ varies dynamically from cell to cell according to Eqs.~(\ref{Eq: r_crit}, \ref{Eq: mcl}, and \ref{Eq: rhocl}), and may exceed the prescribed boundaries $m_\textrm{min}$ and $m_\textrm{max}$. When this occurs, the functions $r^{-1}_\textrm{loss}$, $r^{-1}_\textrm{growth}$, and $r^{-1}_\textrm{drag}$ become ill-defined and can drop rapidly to negative values. To stabilize this behavior, we set these functions to zero whenever they become negative.
The dependence of the $r^{-1}_\textrm{*}$ coefficients on the model parameters is discussed in Appendix~\ref{app: r-1}.

\begin{table}[t]
\caption{Summary of the cloud model parameters.}
\label{tab: model parameters}
\centering
\begin{tabular}{cc}
\hline\hline
 & Interaction model parameters \\
\hline
$T_\textrm{cold}$ & Cold phase temperature \\
$\varepsilon_P$ & Cold to hot pressure ratio \\
$f_\textrm{mix}$ & Mixing strength \\
$f_\textrm{turb}$ & Turbulent to relative velocity ratio \\
\hline
 & Cloud model parameters \\
\hline
$r_\textrm{cl}$ & Cloud size  (delta model) \\
$m_\textrm{min}$ & Minimum cloud mass (power law model) \\
$m_\textrm{max}$ & Maximum cloud mass (power law model) \\
$\gamma$ & Low-mass power law slope (power law model) \\
\hline
\end{tabular}
\tablefoot{In the delta model the cloud mass ($m_\textrm{cl}$), in the power-law model the critical cloud mass and size ($m_\textrm{crit}$ and $r_\textrm{crit}$), and in both models the cloud density ($\rho_\textrm{cl}$), are quantities derived from the macroscopic gas state.}
\end{table}

In both the delta and power law models, the source terms are proportional to the cold-phase mass within the cell, $M_\textrm{cold}$. As a consequence, if $\dot{M}_\textrm{growth} > \dot{M}_\textrm{loss}$, the cold phase will tend to grow increasingly rapidly, eventually draining the hot phase. This behavior reflects an inherent limitation of source terms derived from cloud-crushing simulations, which typically cover only a few cloud-crossing times and do not capture the long-term evolution of clouds embedded in a hot medium \citep[see][for a discussion how to mitigate this problem in the compressible 2-fluid framework]{Das2024}. To mitigate this issue in the case of a pressureless fluid, we rescale the source terms by a sigmoid function that vanishes as $M_\textrm{cold}/M_\textrm{hot} \geq 10^5$. Similarly, we apply the same suppression when $M_\textrm{hot}/M_\textrm{cold} \geq 10^5$ to prevent the cold phase from becoming vanishingly small.
For clarity, we summarize the parameters of our cloud models in Table~\ref{tab: model parameters}.

To trace the fate of initially hot, cold and feedback gas, we employ passive scalars. These are kept track of in the hot and cold phases separately and follow the mass fluxes across interfaces. For the mass exchange between fluids, any given $i$-th scalar, with values $\mathcal{P}_{\textrm{hot},i}$ and $\mathcal{P}_{\textrm{cold},i}$ in the hot and cold phases respectively, the scalar equations take the form
\begin{align}
    &\frac{d}{dt}\bigl(M_\textrm{hot}\mathcal{P}_{\textrm{hot},i} \bigr) = \dot{M}_\textrm{loss} \mathcal{P}_{\textrm{cold},i} - \dot{M}_\textrm{growth} \mathcal{P}_{\textrm{hot},i}, \label{Eq: scalars hot}\\
    &\frac{d}{dt}\bigl(M_\textrm{cold}\mathcal{P}_{\textrm{cold},i} \bigr) = -\dot{M}_\textrm{loss} \mathcal{P}_{\textrm{cold},i} + \dot{M}_\textrm{growth} \mathcal{P}_{\textrm{hot},i},
    \label{Eq: scalars cold}
\end{align}
and are evolved in the source term integrator along with the other exchange terms.

More details about the cloud model are provided in Appendix~\ref{app: model}. A comparison between delta and power law distribution is presented in  Appendix~\ref{app: delta-ppl} and comparisons to similar models in literature in Appendix~\ref{app: literature}.

\section{M82 simulations: setup and suite} \label{sec: M82}

To study the behavior of the model described in Section~\ref{Sec: Model}, we carried out a suite of simulations aimed at reproducing the multiphase galactic outflow of the starburst galaxy M82. Our primary goal is to demonstrate that low-resolution simulations employing two-fluid hydrodynamics and sub-grid modeling of cold cloud–hot phase interactions can effectively reproduce the results of high-resolution simulations that explicitly resolve cold clouds and their coupling with the ambient hot gas. Specifically, we align our setup with the high–resolution simulation of \citet{Schneider20} (hereafter \citetalias{Schneider20}), adopting a similar initial conditions and energy injection prescriptions (Section~\ref{sec: setup}) to ensure a meaningful comparison.

\subsection{Setup} \label{sec: setup}

All simulations are performed with the moving-mesh code \textsc{Arepo} \citep{Springel10, Pakmor16, Weinberger20}, with refinement, derefinement, and mesh-motion criteria modified to meet the specific requirements of our setup, as detailed in Appendix~\ref{app: refinement}.
The simulation box has dimensions of 20 kpc $\times$ 20 kpc $\times$ 20 kpc, with the galactic disc lying in the plane $z = 10$ kpc, and centered in $(x=10\, \textrm{kpc}, y=10\, \textrm{kpc})$.
We initialize a gaseous disc in vertical hydrostatic equilibrium and in dynamical equilibrium with static stellar and dark matter potentials by means of the code \textsc{gd\_basic} by \cite{Lupi15}, following the methodology of \citetalias{Schneider20}. The properties and parameters of the three components are as follows:
\begin{itemize}
    \item \textbf{Gas:} Initially sampled with $N_\textrm{gas} = 2\cdot 10^4$ particles, distributed according to an exponential radial profile with scale radius $R_\textrm{gas} = 1.6$ kpc and total mass $M_\textrm{gas} = 2.5 \cdot 10^9 \mathrm{M}_\odot$. The gas is initialized at a temperature $T = 10^4$ K,
    with a constant mean molecular weight $\mu = 0.6$. 
    Both, for the initial equilibrium and during the simulation, we neglect the self-gravity of the gas. After initialization, the mass in each cell is redistributed so that a fraction $f_\textrm{cold}$ of the total cell mass is assigned to the cold phase, and the cold gas rotational velocity is determined solely by the gravitational potential, hence slightly higher than that of the hot phase, which also experiences pressure support. The cold phase is initialized at a constant temperature $T_\textrm{cold} = 10^4$ K and remains constant throughout each simulation.
    To fill the simulation volume, a background low-density medium composed of both phases is placed above and below the disc. This component becomes dynamically irrelevant once the outflow reaches those regions, and any information about the initial conditions there is quickly erased.

    \item \textbf{Stars:} Included as a static Miyamoto–Nagai potential \citep{MiyamotoNagai75} with total mass $M_\star = 10^{10} \mathrm{M}_\odot$ and characteristic radius and scale height $R_\star = 0.8$ kpc and $z_\star = 0.15$ kpc, respectively.

    \item \textbf{Dark matter:} Modelled via a static Navarro–Frenk–White potential \citep{NFW96}, with total mass $M_\textrm{DM} = 5\cdot10^{10} \mathrm{M}_\odot$, scale radius $R_\textrm{DM} = 5.3$ kpc and concentration parameter $\mathcal{C} = 10$.
\end{itemize}

The starburst is driven by the injection of mass and energy into the hot phase. To mimic clustered supernova (SN) feedback, we follow the procedure similar to \citetalias{Schneider20}, injecting mass and energy into eight spherical regions of radius $30$ pc, randomly positioned within $1$ kpc from the galactic center. Injection occurs in all resolution elements that lie within these spheres, at a mass rate 
$\dot{M}_\textrm{inj} = \eta_M \dot{M}_\star$ and energy rate $\dot{E}_\textrm{inj} = \eta_E \cdot (10^{51}\, \mathrm{erg}/100\, \mathrm{M}_\odot)\cdot \dot{M}_\star$,
where $\dot{M}_\star = 20\, \mathrm{M}_\odot\, \mathrm{yr}^{-1}$, $\eta_M = 0.1$ and $\eta_E = 0.325$.  While the value of $\eta_M$ is the same as that adopted by \citetalias{Schneider20}, they assume a higher energy-loading factor of 1.
Once initialized, the injection spheres rotate in the disc according to the local circular velocity set by the total gravitational potential. Each sphere remains active for $10$ Myr, after which it is replaced by a newly generated one at a different random location. Mass and energy injection begins at $t = 0$ and continues for the entire duration of the simulation, which spans $30$ Myr.
To ensure that the feedback regions are resolved despite the low overall resolution, we implement a refinement/derefinement criterion which enforces a target cell volume of $V_\textrm{disc} = 10^{-5}\, \textrm{kpc}^3$ for gas cells in the disc inner region, such that each injection sphere is resolved by approximately $\gtrsim 6$ cells. Outside the disc, the resolution gradually decreases, with typical cell volumes in the outflow reaching $V_\textrm{crit} = 0.1\, \textrm{kpc}^3$. Further details on the refinement strategy are provided in the Appendix \ref{app: refinement}.

We include radiative cooling in the hot phase as a source term, integrated together with the cold-hot phase interaction terms. The cooling function is the same piecewise parabolic fit to a solar metallicity cooling curve presented in \cite{Schneider18}.

As the outflow reaches the edge of the simulation box, specifically a radial distance $R = 9$ kpc from the center of the box, we apply a boundary sink term. 
Cells are coarsened to a volume $V_\mathrm{out} = 1 \,\mathrm{kpc}^3$, their velocities (in both phases) are damped, the hot-phase temperature is reset to $10^4 \,\mathrm{K}$, the hot-phase density is set to $5.5 \times 10^{-36} \,\mathrm{g\,cm^{-3}}$, and the cold-phase density is set to a value 100 times smaller. This approach causes the outflow to gradually disappear at this radius, and is de-facto an outflow boundary condition for the developing (potentially supersonic) flow.

In addition, to trace the origin of cold gas in the outflow, we introduce three passive scalar fields for each phase, that we indicate with $\mathcal{P}_{\textrm{phase},i}$, where the subscript \emph{phase} indicates the phase it belongs to and $i$ the scalar type. These three types of scalars track the fractional mass contributions to the given phase from:  the initial hot phase ($i=0$),  the initial cold phase ($i=1$), and the SN-injected material ($i=2$). At initialization, hot and cold gas phases are assigned scalar values of $(1, 0, 0)$ and $(0, 1, 0)$, respectively. SN-injected material carries $(0, 0, 1)$. Whenever mass is exchanged, either between phases within a cell (due to mixing, Eqs. \ref{Eq: scalars hot} and \ref{Eq: scalars cold}), or between cells (due to advection, see \citealt{Weinberger23}), the passive scalars are exchanged proportionally along with the corresponding mass fluxes, which allows us to study the origin of the cold gas in the outflow.

\subsection{Simulation suite} \label{sec: suite}
\begin{table*}[t]
    \caption{Summary of the parameters used in our simulation suite. The following values are kept constant across all runs: $T_\textrm{cold} = 10^4$~K, $\varepsilon_P = 0.143$, $f_\textrm{turb} = 0.1$, $m_\textrm{min} = 10^{-2} \, \mathrm{M}_\odot$, $m_\textrm{max} = 10^{9} \, \mathrm{M}_\odot$ and $\gamma = 0$.}
    \centering
    \begin{tabular}{l|cccccc}
        \hline\hline
        \vspace{0.001\linewidth}
        name & \shortstack{clouds \\ distribution} & \shortstack{drag \\ source term} & $f_\textrm{mix}$ & $V_\textrm{crit}$ [kpc$^3$] & $f_\textrm{cold}$ & $r_\textrm{cl}$ [pc] \\
        \hline
        {\labsim{fid}} & power law  & \ding{51}  &  0.003 & 0.1 & 0.65 & \ding{55}\\
        {\labsim{NoSources}} & power law  & \ding{55}  & 0 & 0.1& 0.65 &  \ding{55}\\
        {\labsim{NoDrag\_fmix0.03}} & power law  & \ding{55}  &  0.03& 0.1 & 0.65  &  \ding{55}\\
        \hline
        {\labsim{fmix0.03}} & power law   & \ding{51} & 0.03 & 0.1 & 0.65 &  \ding{55}\\
        {\labsim{fmix0.1}} & power law  & \ding{51}  & 0.1 & 0.1 & 0.65 & \ding{55}\\
        {\labsim{fmix0.3}} & power law  & \ding{51}  & 0.3 & 0.1 & 0.65 &\ding{55}\\
        {\labsim{fmix1}} & power law  & \ding{51}  & 1 & 0.1 & 0.65 &\ding{55}\\
        \hline
        {\labsim{Vc0.001}} & power law  & \ding{51}  & 0.003  & 0.001 & 0.65& \ding{55}\\
        {\labsim{Vc0.01}} & power law  & \ding{51}  & 0.003 & 0.01 & 0.65 &\ding{55}\\
        {\labsim{Vc0.03}} & power law  & \ding{51}  & 0.003 & 0.03 & 0.65 &\ding{55}\\
        {\labsim{Vc0.3}} & power law  &\ding{51}   & 0.003 & 0.3 & 0.65 &\ding{55}\\
        {\labsim{Vc1}} & power law  & \ding{51}  & 0.003 & 1 & 0.65 &\ding{55}\\
        \hline
        {\labsim{fcld0.1}} & power law  & \ding{51}  & 0.003 & 0.1  &0.1 &\ding{55}\\
        {\labsim{fcld0.9}} & power law  & \ding{51}  & 0.003 & 0.1 & 0.9&\ding{55}\\
        {\labsim{rcl0.01}} & delta & \ding{51}  & 0.003 & 0.1 & 0.65 & 10\\
        {\labsim{rcl0.6}} & delta & \ding{51}  & 0.003 & 0.1 & 0.65 & 600\\
        \hline
    \end{tabular}
    \label{tab: sim parameters}
\end{table*}
We performed a suite of simulations, summarized in Table~\ref{tab: sim parameters}, with the following characteristics:
\begin{itemize}
    \item The fiducial run ({\labsim{fid}}) was tuned by adjusting a subset of model and setup parameters so that the resulting outflow properties reproduce those of the simulation by \citetalias{Schneider20}, as discussed in Section~\ref{sec: schneider}. The run {\labsim{NoSources}} is identical to {\labsim{fid}} except that all source terms are disabled, while {\labsim{NoDrag\_fmix0.03}} switches off the drag force and employs a mixing coefficient ten times larger than the fiducial value.
    \item The runs {\labsim{fmix*}} are identical to {\labsim{fid}} except for an increased mixing coefficient. These are presented in Section~\ref{sec: fmix} to explore how enhanced phase mixing affects the properties of the multiphase outflow.
    \item All simulations discussed in the main text are performed at the same resolution. In Appendix~\ref{sec: resolution}, we show that the results are robust even under extreme resolution changes. The {\labsim{Vc*}} runs differ from {\labsim{fid}} only in the value of the characteristic cell volume $V_\mathrm{crit}$, which controls mesh refinement and derefinement in the outflow region, i.e. the numerical resolution. 
    \item The {\labsim{fcld*}} runs investigate the dependence of the model on the initial cold gas mass fraction in the disc, using both higher and lower $f_\mathrm{cold}$ values relative to {\labsim{fid}}. Finally, the {\labsim{rcl*}} runs compare the power-law cloud model with the delta-function cloud model for a couple representative cloud sizes. Both sets of runs are discussed in Appendix~\ref{sec: other}.
\end{itemize}

\section{M82 simulations: results} \label{sec: results}
\begin{figure*}
    \centering
    \includegraphics[width=1\linewidth]{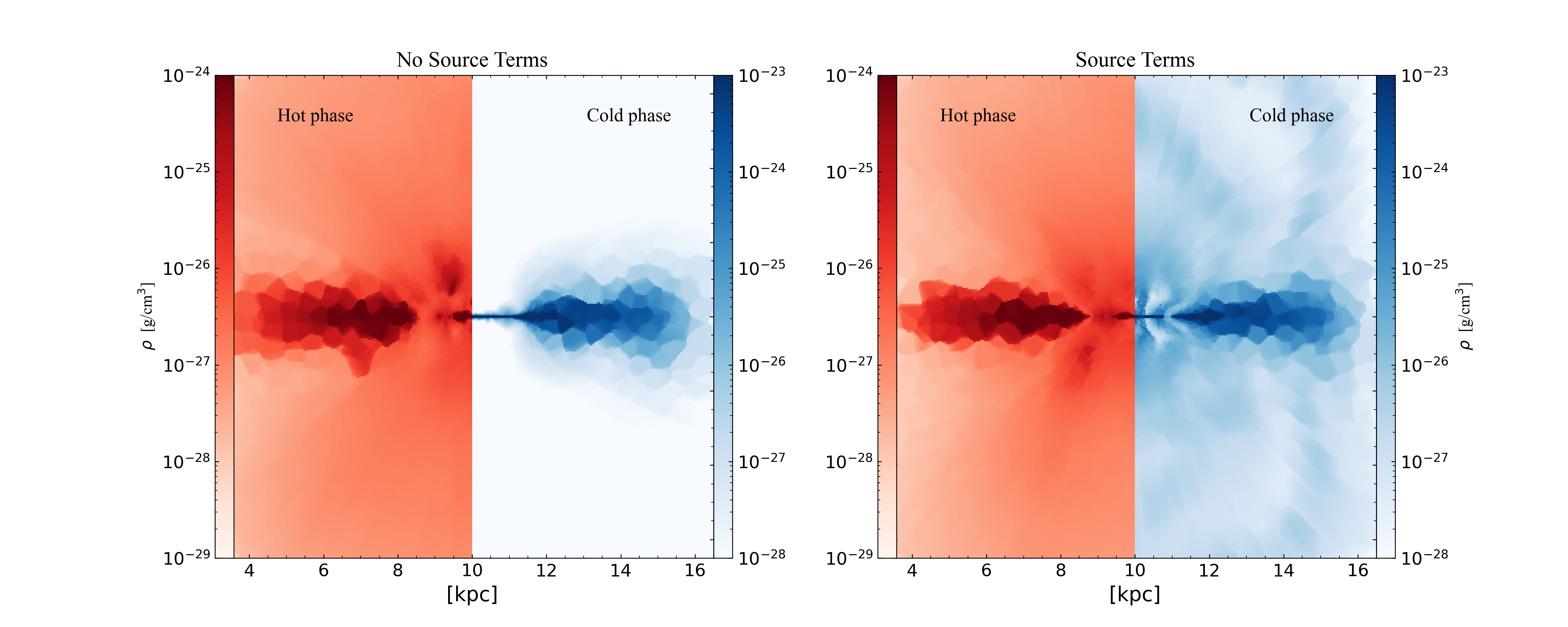}
    \caption{Density slices of the {\labsim{NoSources}} run (left panel) and the {\labsim{fid}} run (right panel). The density in each cell is defined as the mass in the respective phase divided by the volume of the cell. The slices show time averages over snapshots taken in the interval $T = (27, 30)$~Myr. In each panel, the left side corresponds to the hot phase and the right side to the cold phase.}
    \label{fig: Maps}
\end{figure*}
To illustrate the overall behavior of the simulation, Figure~\ref{fig: Maps} shows density slices for the {\labsim{NoSources}} run (left panel) and the {\labsim{fid}} run (right panel). In both panels, the left side of the box displays the density of the hot phase, while the right side shows that of the cold phase.  Each panel presents a time average over snapshots taken in the interval $T = (27, 30)$ Myr after the onset of SNe injection. 
In both cases, the injection of mass and energy into the hot phase drives an hot outflow that fills the entire simulation volume. However, when the source terms are switched off (left panel), no cold outflow develops and all cold gas remains confined to the disc. In contrast, when the source terms are active (right panel), mass, momentum, and energy are exchanged between phases, leading to the emergence of a cold outflow from the cold gas disk.

\subsection{Comparison with a high-resolution M82 simulation} \label{sec: schneider}

\begin{figure}
    \centering
    \includegraphics[width=1\linewidth]{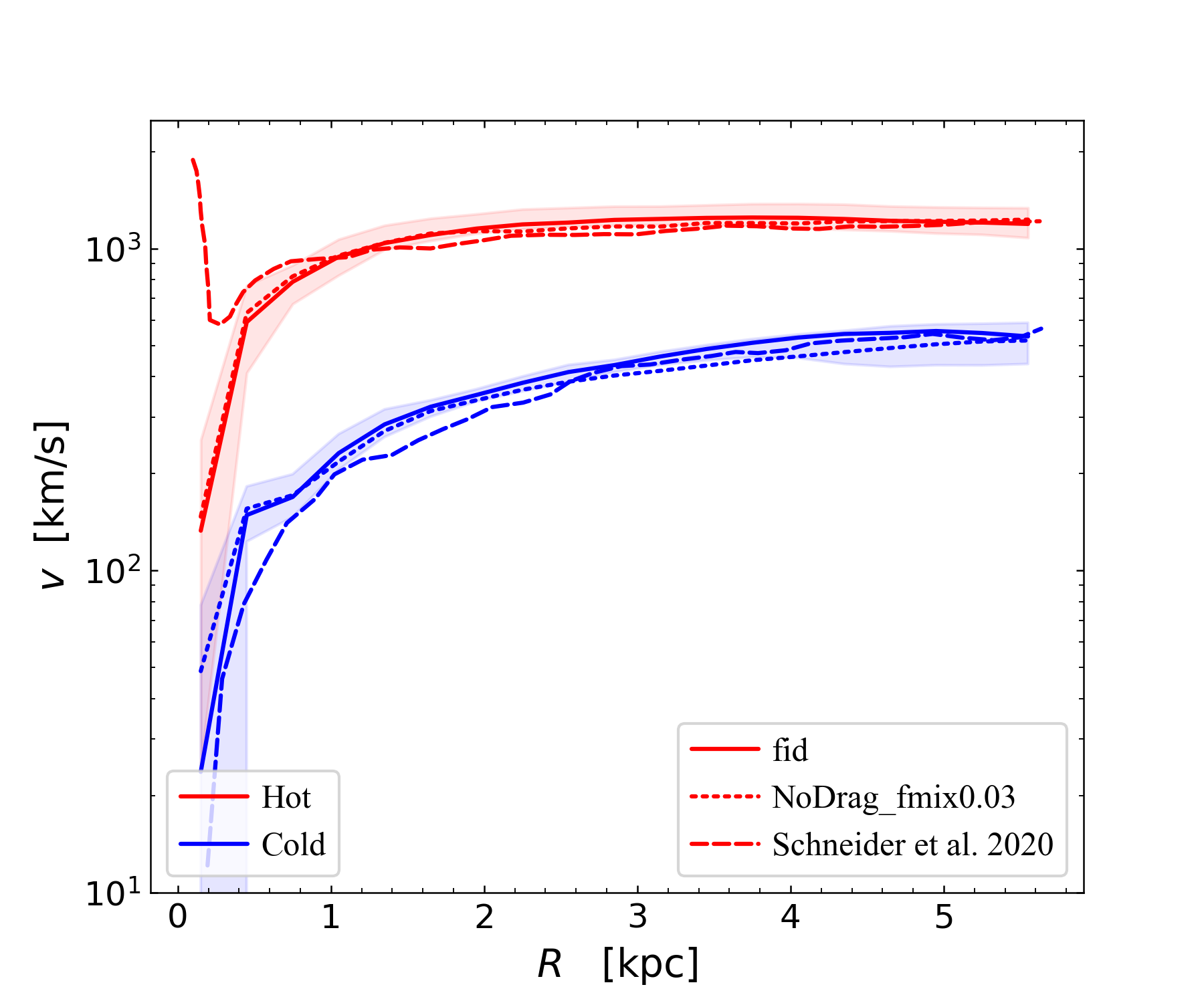}
    \caption{Radial velocity profiles of the hot (red) and cold (blue) phases from our {\labsim{fid}} (solid lines) and {\labsim{NoDrag\_fmix0.03}} (dotted lines) simulations, compared with the results of \citetalias{Schneider20} (dashed lines). All profiles show medians computed within a cone of semi-opening angle $30^\circ$. Shaded regions indicate the 16th and 84th percentiles for the {\labsim{fid}} run.
}
    \label{fig: Schneider velocity}
\end{figure}

\begin{figure}
    \centering
    \includegraphics[width=1\linewidth]{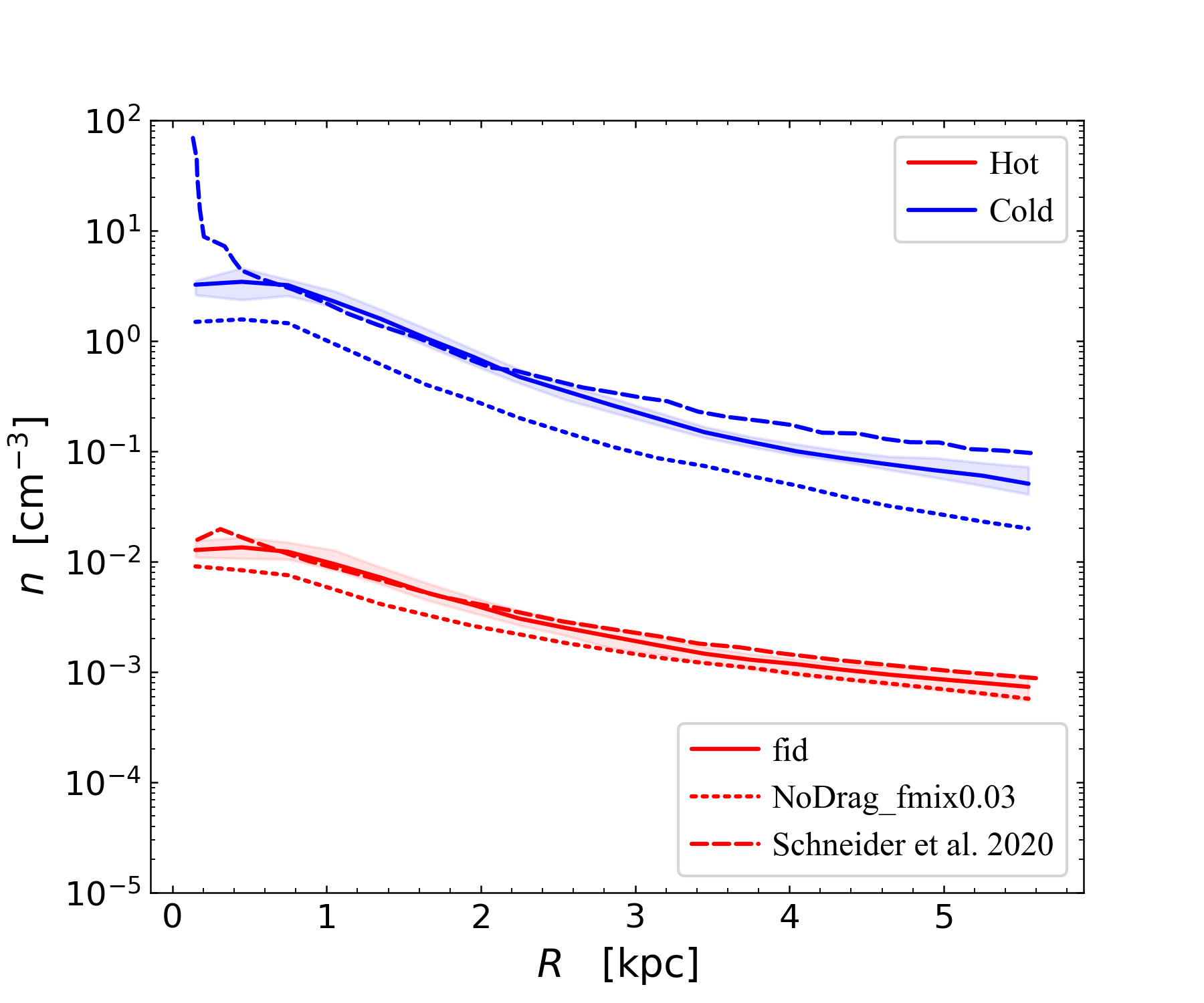}
    \caption{Radial density profiles of the hot (red) and cold (blue) phases from our {\labsim{fid}} (solid lines) and {\labsim{NoDrag\_fmix0.03}} (dotted lines) simulations, compared with the results of \citetalias{Schneider20} (dashed lines). All profiles show medians computed within a cone of semi-opening angle $30^\circ$. Shaded regions indicate the 16th and 84th percentiles for the {\labsim{fid}} run.}
    \label{fig: Schneider density}
\end{figure}

\begin{figure}
    \centering
    \includegraphics[width=1\linewidth]{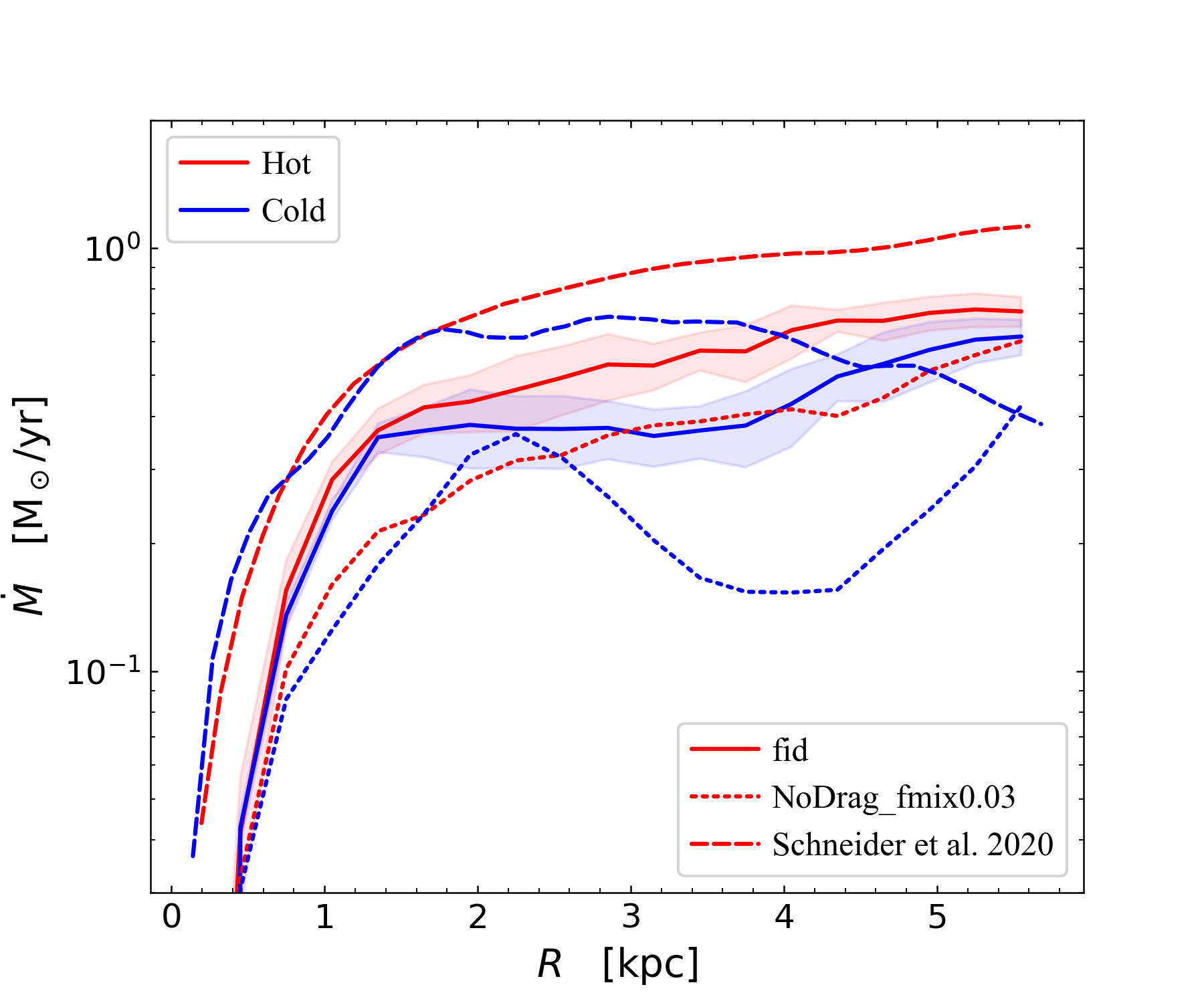}
    \caption{Radial mass outflow rate profiles of the hot (red) and cold (blue) phases from our {\labsim{fid}} (solid lines) and {\labsim{NoDrag\_fmix0.03}} (dotted lines) simulations, compared with the results of \citetalias{Schneider20} (dashed lines). All profiles are computed by summing the mass outflow rate from each cell at a given radius within a cone of semi-opening angle $30^\circ$. Shaded regions indicate the time variability in the interval $(27, 30)$~Myr for the {\labsim{fid}} run.
}
    \label{fig: Schneider Mdots}
\end{figure}

In this section, we compare two of the simulations listed in Section~\ref{sec: suite} with the high-resolution simulation of \citetalias{Schneider20}. In our fiducial simulation, the parameters 
$\mathcal{U} = \{ f_\textrm{cold}, \eta_\textrm{E}, f_\textrm{mix}, m_\textrm{min}, m_\textrm{max}, \gamma, \varepsilon_P \}$ 
are chosen such that the macroscopic properties of the resulting outflow, namely the density, velocity, and mass outflow rate of each phase, match those measured in the high-resolution simulation of \citetalias{Schneider20}. The fiducial parameter values are listed in Table~\ref{tab: sim parameters}.
Figures~\ref{fig: Schneider velocity}, \ref{fig: Schneider density}, and \ref{fig: Schneider Mdots} compare the macroscopic outflow properties in our {\labsim{fid}} simulation (solid lines with shaded regions) to those from \citetalias{Schneider20} (dashed lines, extracted from their Figs.~6--8). For reference, we also show the {\labsim{NoDrag\_fmix0.03}} simulation (dotted lines), which is identical to {\labsim{fid}} except that the drag force is disabled and the mixing strength is increased to $f_\textrm{mix} = 0.03$, one order of magnitude higher than the fiducial value.
In all figures, red lines correspond to the hot phase and blue lines to the cold phase.

For the velocities and densities, the profiles are computed as follows. First, we calculate the time-averaged density and velocity structures over the interval $T$. Then, for each quantity, its value at radius $r$ is given by the median within the shell $(r,\, r+0.3 \,\textrm{kpc})$ inside a cone with semi-opening angle $\theta = 30^\circ$, considering both upward and downward outflows. In our fiducial run, the boundaries of the shaded regions correspond to the 16th and 84th percentiles, which reflect the angular variability within each radial bin. 
The same procedure is used by \citetalias{Schneider20}, except they omit the initial time-averaging step, which we employ to reduce shot noise arising from our significantly lower resolution, and instead use a single snapshot taken at $t = 30$ Myr after the onset of SNe injection.

For the mass outflow rates, we proceed as follows. For each snapshot in the interval $T$, we compute the mass outflow rate in the radial bin  $(r,\, r+0.3 \,\textrm{kpc})$ as 
$\sum_i m_i \, v_i/\Delta r$ considering all cells within the cone of semi-aperture $\theta = 30^\circ$, as above. The mass outflow rate at $r$ is then taken as the mean across the considered snapshots, and the boundaries of the shaded regions correspond to the standard deviation. Again, \citetalias{Schneider20} adopt the same method, but only for the snapshot at $t = 30$ Myr, and therefore without time-averaging.

Overall, we find that the macroscopic properties of the outflow in our {\labsim{fid}} simulation are in very good agreement with those from the high-resolution simulation of \citetalias{Schneider20}. Even for the mass outflow rates, where the mismatch appears larger, the differences remain within a factor of $\lesssim 2$. Regarding the discrepancy in the cloud densities, \citetalias{Schneider20} note that their profile may be affected by resolution effects, which cause the pressure equilibrium between phases to vary with radius. In our model the pressure ratio between phases is assumed to be constant everywhere. 
What makes the overall agreement noteworthy is that, thanks to our different discretization of the problem, i.e., two-fluid hydrodynamics coupled with subgrid models for the interaction between the resolved hot phase and unresolved cold clouds, we are able to obtain quantitatively similar results to \citetalias{Schneider20} for the properties of the multiphase outflow, despite employing a significantly lower resolution.
More specifically, within the outflow, the typical cell volume in our simulations is $V_\textrm{crit} = 0.1 \,\textrm{kpc}^3$, corresponding to a mean linear cell size of $l_\textrm{cell} \simeq 460$~pc, whereas in \citetalias{Schneider20} the cell size is $5$~pc. This represents a factor of $\sim 100$ in linear resolution, corresponding to a factor of $\sim 10^6$ in the number of resolution elements.

In Figures~\ref{fig: Schneider velocity}, \ref{fig: Schneider density}, and \ref{fig: Schneider Mdots}, the purpose of showing the {\labsim{NoDrag\_fmix0.03}} simulation is to highlight that there is a degeneracy between drag and mixing in launching the cold outflow. 
In our subgrid treatment of the drag force, we assume that the total drag force on the cold phase in a cell is given by the sum of the drag forces acting on all individual clouds comprising the cold phase. This approach may overestimate the total drag force, as it neglects physical shielding effects within the cloud population (e.g. \citealt{Villares24}). To explore the extreme of this regime, we consider the case of a vanishing drag force and investigate whether mixing alone can accelerate the cold outflow to reproduce the velocity curves of \citetalias{Schneider20}. Figure~\ref{fig: Schneider velocity} shows that, by increasing $f_\textrm{mix}$ by an order of magnitude above the fiducial value, and in the absence of drag, we can still recover the same velocity structure of the multiphase outflow. 
This demonstrates that the set of parameter values of {\labsim{fid}} used to match \citetalias{Schneider20} is not unique, and that different combinations of drag and mixing strengths can reproduce the same velocity structure reported in \citetalias{Schneider20}\footnote{This is in agreement with \citetalias{Schneider20} (their section 4.6), as they demonstrate that ram pressure cannot have a dominant effect on the acceleration of the cold gas and some degree of mixing acceleration is required to explain the observed cold velocity profile.}. The match between {\labsim{NoDrag\_fmix0.03}} and \citetalias{Schneider20} in densities and mass outflow rates is poorer than in {\labsim{fid}}. We suspect that this is a consequence of  only modifying the parameter $f_\textrm{mix}$ to match velocities, without a full recalibration of all the parameters $\mathcal{U}$ to reproduce all observables. Nevertheless, even in densities and mass outflow rates, the {\labsim{NoDrag\_fmix0.03}} predictions remain within an order of magnitude of \citetalias{Schneider20}. The most notable discrepancy is in the cold cloud density. This is a consequence, as shown in Appendix~\ref{app: fmix}, of most of the heating of the hot wind originating from dissipation of drag forces. Without drag, the hot phase is cooler and less pressurized, leading to a lower cloud densities (Eq.~\ref{Eq: rhocl}).


In the next section, we systematically explore the effects of progressively increasing the mixing efficiency $f_\textrm{mix}$ on the outflow.

\subsection{Variation of the mixing efficiency} \label{sec: fmix}
\begin{figure*}
    \centering
    \includegraphics[width=1.05\linewidth]{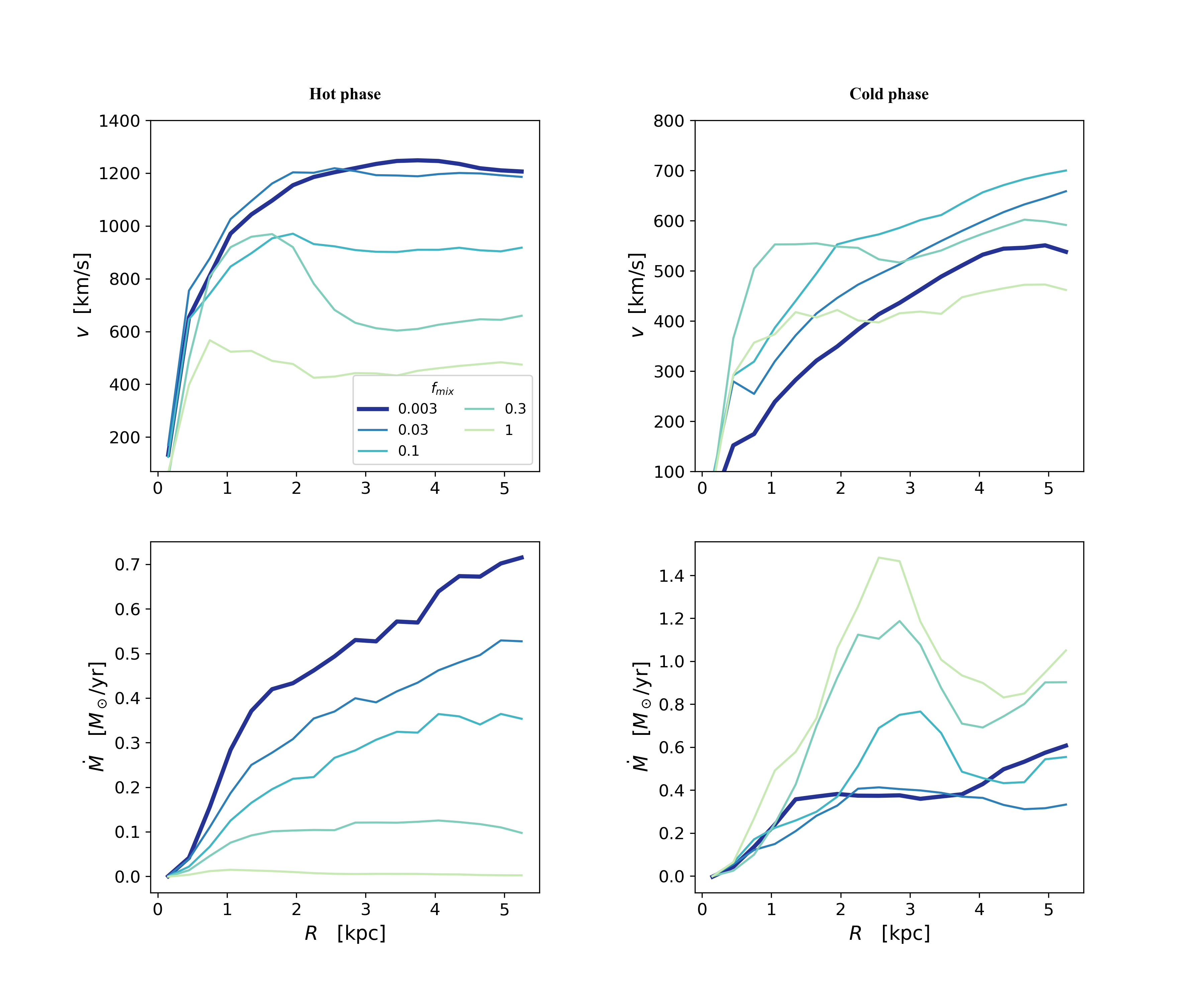}
    \caption{ Velocities and mass outflow rates for the fiducial and {\labsim{fmix*}} runs.  
\emph{Top row:} Radial profiles of the velocity for the hot (left) and cold (right) phases. The velocity at each radius is computed as the median within a radial bin of 0.3 kpc and inside a cone with a semi-opening angle of $30^\circ$.  
\emph{Bottom row:} Radial profiles of the mass outflow rate for the hot (left) and cold (right) phases. The mass outflow rate at each radius is the sum of the outflow rates of all cells within a 0.3 kpc radial bin and inside the same $30^\circ$ cone. 
}
    \label{fig: mix_vm}
\end{figure*}
In this section we explore the effects of enhancing mass exchange between the cold and hot phases by increasing the parameter $f_\textrm{mix}$ above its fiducial value of $0.003$. The {\labsim{fmix\_*}} simulations, which differ from the fiducial run {\labsim{fid}} only in having larger values of $f_\textrm{mix}$ (see Table~\ref{tab: sim parameters}), provide a basis for assessing these changes. Figure~\ref{fig: mix_vm} shows the velocities (top row) and mass outflow rates (bottom row) of the hot and cold phases (left and right columns, respectively) for the {\labsim{fid}} and {\labsim{fmix*}} simulations. The fiducial run is highlighted with thicker lines for reference. Velocity and mass outflow rate profiles are computed as outlined in the previous section. 

Figure~\ref{fig: mix_vm} reveals a couple of interesting trends:
i) As $f_\textrm{mix}$ increases, the velocity of the hot phase decreases, while the cold-phase velocity shows a non-monotonic behavior. Specifically, it increases with $f_\textrm{mix}$ up to 0.1, but then decreases for higher values. Despite this decline, the velocity profile for $f_\textrm{mix} \geq 0.1$ still shows a steeper rise at $R \lesssim 1$~kpc compared to runs with lower $f_\textrm{mix}$;
ii) At higher $f_\textrm{mix}$ values, the hot-phase mass outflow rate decreases, whereas the cold-phase outflow rate increases. 

To interpret the velocity behavior, we analyze the accelerations acting on the two phases, separating hydrodynamical from source–term contributions. The hydrodynamical acceleration for the hot phase, the source–term accelerations for the hot and for cold phases are
\begin{align}
    a_\textrm{hydro} &= -\frac{\partial_r P_\textrm{hot}}{\rho_\textrm{hot}}, \\
    a_\textrm{h,sources} &= -\frac{|\mathbf{\dot{P}_\textrm{drag}}|}{M_\textrm{hot}} - \frac{\dot{M}_\textrm{loss}}{M_\textrm{hot}}\,v_\textrm{rel}, 
    \label{Eq: acceleration hot}\\
    a_\textrm{c,sources} &= +\frac{|\mathbf{\dot{P}_\textrm{drag}}|}{M_\textrm{cold}} + \frac{\dot{M}_\textrm{growth}}{M_\textrm{cold}}\,v_\textrm{rel}, 
    \label{Eq: acceleration cold}
\end{align}
respectively, where
the second terms in the source equations are obtained by combining Eqs.~(\ref{Eq: pdot pop mix}) and (\ref{Eq: ppl Mdot}) and $v_\textrm{rel}$ is the difference between the hot and cold radial velocities. 
\begin{figure*}
    \centering
    \includegraphics[width=1.05\linewidth]{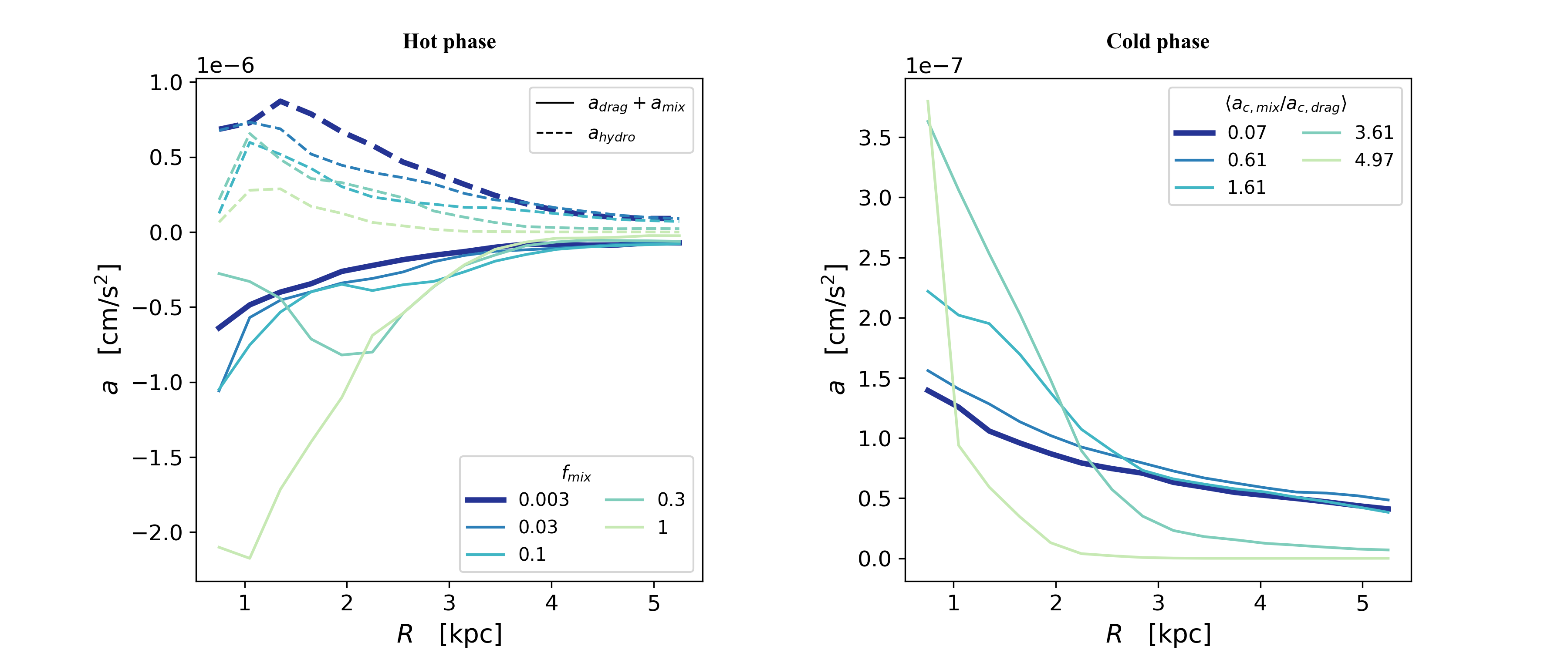}
    \caption{Source-term accelerations (solid lines) for the hot (left) and cold (right) phases in the fiducial and {\labsim{fmix*}} runs. The pressure gradient accelerations of the hot phase are shown as dashed lines in the left panel. All curves represent medians computed in the same way as the velocities in Fig.~\ref{fig: mix_vm}. The legend in the right panel indicates the radial average of the ratio between the mixing and drag contributions to the cold phase acceleration.}
    \label{fig: accelerations}
\end{figure*}
Figure~\ref{fig: accelerations} shows the source–term accelerations (solid lines) for the hot phase (left panel) and the cold phase (right panel) for different $f_\textrm{mix}$ values. In the left panel, the dashed lines indicate the hot–phase pressure–gradient acceleration. The legend in the right panel lists the values of $\langle a_\textrm{c,mix} / a_\textrm{c,drag} \rangle$, the radial average of the ratios between the second and the first term in Eq.~(\ref{Eq: acceleration cold}).

For the hot phase, two effects combine to produce progressively lower accelerations as $f_\textrm{mix}$ increases. First, a higher $f_\textrm{mix}$ enhances the source–term deceleration due to mixing by increasing $\dot{M}_\textrm{loss}$. Second, it reduces the pressure–gradient acceleration driving the hot wind. The latter effect is more subtle and is discussed quantitatively in Appendix~\ref{app: fmix}. In brief, stronger mixing increases the deceleration of the hot phase and the acceleration of the cold phase, bringing their velocities progressively closer and thus reducing $v_\textrm{rel}$. Since the drag–heating rate $\dot{u}_\textrm{drag}$ (Eq.~\ref{Eq: dotu pop drag}) scales as $v_\textrm{rel}^2$, it drops rapidly for higher mixing. As $\dot{u}_\textrm{drag}$ dominates the hot–phase heating (i.e. it is the main term in Eq.~\ref{Eq: ppl udot}), its decline leads to a lower hot–phase temperature and, consequently, a weaker pressure gradient. This process shows that the interaction with the cold phase alters the hot–phase thermodynamics in a non-trivial way that ultimately affects its kinematics.

For the cold phase, with increasing $f_\textrm{mix}$, the mixing–driven acceleration eventually dominates over drag (for $f_\textrm{mix} \gtrsim 0.1$; see legend in the right panel) as the coefficient $\dot{M}_\textrm{growth}$ in Eq.~\ref{Eq: acceleration cold} increases. At high mixing, this additional acceleration is most effective near the wind base, leading to a rapid initial boost of the cold phase toward the hot–phase velocity. This faster convergence reduces $v_\textrm{rel}$ more quickly, causing a steeper drop in the total source–term acceleration, since both terms in Eq.~\ref{Eq: acceleration cold} scale with $v_\textrm{rel}$.  
For $f_\textrm{mix} \lesssim 0.1$, the deceleration of the hot phase and the acceleration of the cold phase both increase with mixing, but only moderately; the cold phase consistently lags behind the hot phase throughout the simulated domain. In contrast, for $f_\textrm{mix} > 0.1$, the cold phase undergoes a stronger initial acceleration, but this is abruptly halted around $R \sim 1$--$2$~kpc due to the sharp decline in $v_\textrm{hot}$. The sustained high values of mixing then enforce a tighter coupling between the phases, causing them to expand at approximately the same velocity.

\begin{figure}
    \centering
    \includegraphics[width=1\linewidth]{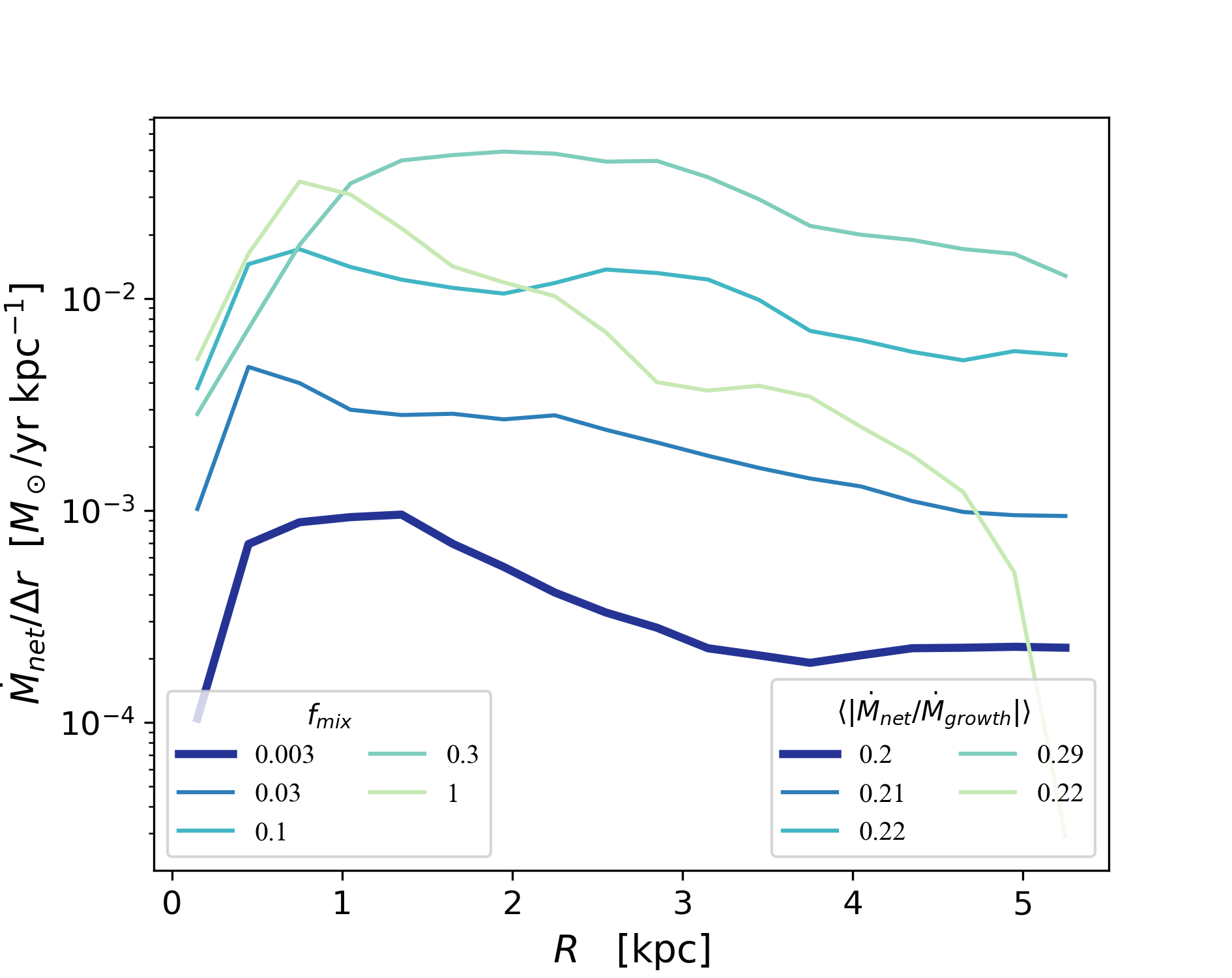}
    \caption{Radial profiles of $\dot{M}_\textrm{net}$, representing the total net mass transfer from the hot phase to the cold phase for the fiducial and {\labsim{fmix*}} runs. $\dot{M}_\textrm{net}$ is calculated in each radial bin of $\Delta r =0.3$ kpc as the sum of $\dot{M}_\textrm{growth} - \dot{M}_\textrm{loss}$ over all cells within a cone of semi-opening angle $30^\circ$. The legend in the bottom right displays the radial average ratios of $\dot{M}_\textrm{net}$ to $\dot{M}_\textrm{growth}.$ 
}
    \label{fig: Mnet}
\end{figure}
Regarding the mass outflow rates, the progressive decline of the hot–phase outflow rate with increasing mixing arises from two effects: the reduction in the hot–phase velocity (top left panel in Fig. \ref{fig: mix_vm}) and the leakage of mass from the hot to the cold phase. For the latter, Figure~\ref{fig: Mnet} shows the radial profiles of $\dot{M}_\textrm{net}$, computed in each radial bin as the sum of $\dot{M}_\textrm{growth} - \dot{M}_\textrm{loss}$ over all cells within a cone of semi–opening angle $30^\circ$. In all simulations, there is a net mass transfer from the hot to the cold phase, and this transfer grows with increasing $f_\textrm{mix}$, except for the most extreme case ($f_\textrm{mix} = 1$). In that case, $\dot{M}_\textrm{net}$ is initially larger than in simulations with smaller mixing for $R \lesssim 1$~kpc, but then declines at larger radii. 
The mass leakage from the hot phase into the cold phase, together with the increase in cold–phase velocity (top–right panel of Fig.~\ref{fig: mix_vm}), both contribute to the systematically higher cold–phase mass outflow rates observed with increasing $f_\textrm{mix}$. However, for $f_\textrm{mix} \gtrsim 0.3$, the trend appears to break: the cold mass outflow rate remains high despite the decrease in velocity and the reduction in $\dot{M}_\textrm{net}$ seen for $f_\textrm{mix} = 1$.  This apparent inconsistency arises because Figure~\ref{fig: Mnet} only indicates the local direction and magnitude of the mass exchange between phases, and cannot be directly translated into the instantaneous mass outflow rate at a given radius. The latter depends not only on the current $\dot{M}_\textrm{net}$ at that radius, but also on the integrated history of material that has been transferred and accelerated at smaller radii at earlier times.

\subsection{The origin of the cold phase in the outflow} \label{sec: origin}
\begin{figure*}
    \centering
    \includegraphics[width=1\linewidth]{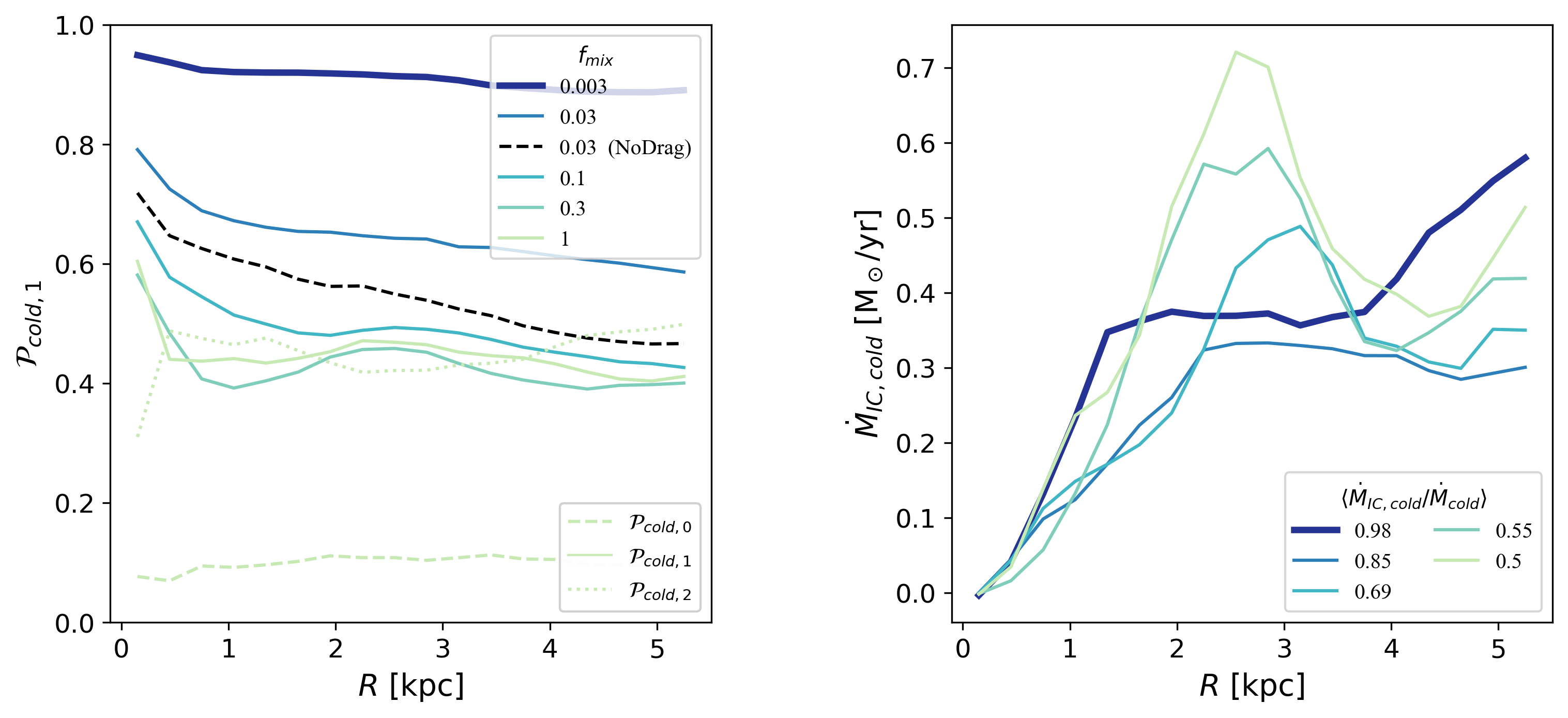}
    \caption{ (\emph{Left panel}) Radial profiles of the fraction of cold mass originating from cold gas in the initial conditions, for the fiducial (thicker line), {\labsim{NoDrag\_fmix0.03}} (dashed line) and {\labsim{fmix*}} simulations. Each curve shows the median within radial bins of $\Delta r = 0.3$~kpc and within a cone of semi–opening angle $30^\circ$. For the {\labsim{fmix1}} run, the dashed and dotted lines indicate the fractions of cold mass originating from hot gas in the initial conditions and from SNe, respectively. (\emph{Right panel}) Radial profiles of the mass outflow rate of gas that was \emph{initially} cold, computed following Eq.~(\ref{Eq:Mdot_IC}). The legend reports the radial averages of the ratio between this quantity and the cold mass outflow rate (see Fig.~\ref{fig: mix_vm}, bottom-left panel).
 }
    \label{fig: mix scalars}
\end{figure*}
The mass and energy injection to drive an outflow is performed only in the hot phase, while the cold phase is initially present solely in the disc. Nevertheless, in all simulations with source terms enabled, a cold-phase outflow is also present. It is therefore natural to ask where the cold phase in the outflow originates from, i.e., if it is dominated by originally cold material, or if condensation out of the hot phase dominates the mass budget.  
To address this, we examine the evolution of the scalar $\mathcal{P}_\textrm{cold,1}$, which traces the fraction of cold-phase mass that was already in the cold phase in the initial condition. Before analyzing the results from our simulations, we first consider a simplified toy model to gain intuition for how $\mathcal{P}_\textrm{cold,1}$ is expected to evolve.
We assume that in a given cell: i) $\dot{M}_\textrm{loss} = \dot{M}_\textrm{growth} \equiv \dot{M}$, as suggested by $(\dot{M}_\textrm{growth} - \dot{M}_\textrm{loss}) / \dot{M}_\textrm{growth} \sim 0.2$ (see legend in Fig.~\ref{fig: Mnet}), and ii) the masses of the phases in the cell remain constant in time. Under these assumptions, Eqs.~(\ref{Eq: scalars hot}) and (\ref{Eq: scalars cold}) give
\begin{equation}
    \mathcal{P}_\textrm{cold,1}(t) = \zeta \bigl(1-e^{-t/\tau}\bigr) + e^{-t/\tau},
    \label{Eq: pcold}
\end{equation}
where $\zeta = M_\textrm{cold} / (M_\textrm{cold} + M_\textrm{hot})$ and $\tau = \zeta (1-\zeta) (M_\textrm{cold} + M_\textrm{hot}) / \dot{M}$. At $t=0$, $\mathcal{P}_\textrm{cold,1} = 1$ and over time it tends toward $\zeta$, doing so increasingly faster for higher $\dot{M}$.
Therefore, in the low-mixing regime, Eq.~(\ref{Eq: pcold}) implies that $\mathcal{P}_\textrm{cold,1} \lesssim 1$, whereas when mixing is stronger, it pushes $\mathcal{P}_\textrm{cold,1}$ toward $\zeta$. This also implies a radial decline of $\mathcal{P}_\textrm{cold,1}$ within the outflow, since cold gas at the base has had less time to mix compared to material at larger radii.
In practice, $\zeta$ differs from the initial cold-phase mass fraction $f_\textrm{cold}$ of our simulations, because the phase masses evolve due to continuous hot-gas injection by SNe, a net hot-to-cold mass flux of $\sim 0.2 \dot{M}_\textrm{growth}$, and hydrodynamic interactions with neighboring cells. The value of $\zeta$ in the simulations therefore results from the combined effect of these processes.

Turning to the simulations, Fig.~\ref{fig: mix scalars} (left panel) shows radial profiles of $\mathcal{P}_\textrm{cold,1}$ for the fiducial and {\labsim{fmix*}} runs. In the fiducial case, where mixing is minimal, $\mathcal{P}_\textrm{cold,1} \lesssim 1$, indicating that the cold phase experiences little mixing while outflowing. As $f_\textrm{mix}$ increases, $\mathcal{P}_\textrm{cold,1}$ decreases, reflecting progressively stronger mixing. In all runs, $\mathcal{P}_\textrm{cold,1}$ declines with radius, showing that mixing increases with outflow travel distance. Most of this decline takes place within $R \lesssim 1$~kpc, indicating that mixing is strongest in the inner outflow region.
For $f_\textrm{mix} > 0.1$, $\mathcal{P}_\textrm{cold,1}$ does not fall below $\sim 0.4$, suggesting that the asymptotic limit $\zeta$ has been reached. Figure~\ref{fig: mix scalars} (left panel) also shows $\mathcal{P}_\textrm{cold,0}$ and $\mathcal{P}_\textrm{cold,2}$ for the {\labsim{fmix1}} run. These indicate that most mixing of the cold phase occurs with SNe-injected material rather than with hot-phase gas from the initial conditions. The same trend holds for other runs, with $\mathcal{P}_\textrm{cold,2}$ decreasing as $f_\textrm{mix}$ decreases, and $\mathcal{P}_\textrm{cold,0}$ remaining $\lesssim 0.1$ in all cases.

For the hot phase, when mixing is minimal (as in {\labsim{fid}}), most of the material originates from SNe ejecta, with $\mathcal{P}_\mathrm{hot,2} \sim 0.8$. As mixing increases, the hot wind becomes increasingly composed of gas that was initially cold and subsequently stripped from the cold phase and incorporated into the hot outflow. By $f_\mathrm{mix} = 1$, this trend leads to $\mathcal{P}_\mathrm{hot,2} \sim 0.4 -0.6$ and $\mathcal{P}_\mathrm{hot,1} \sim 0.3-0.5$. For all mixing rates, $\mathcal{P}_\mathrm{hot,0} \sim 0.1$.

Figure~\ref{fig: mix scalars} (right panel) shows the radial profiles of the mass outflow rate of gas that was initially in the cold phase. At each radius $r$, this quantity is defined as  
\begin{equation}
    \dot{M}_\mathrm{IC,cold} = \frac{1}{\Delta r} \sum_i \left( 
        \mathcal{P}_\mathrm{hot,1} \, M_\mathrm{hot} \, v_{\mathrm{hot},r} \;+\; 
        \mathcal{P}_\mathrm{cold,1} \, M_\mathrm{cold} \, v_{\mathrm{cold},r} 
    \right)_i ,
    \label{Eq:Mdot_IC}
\end{equation}
where $v_{\mathrm{hot},r}$ and $v_{\mathrm{cold},r}$ are the radial velocities of the hot and cold phases in the $i$-th cell. The sum is performed over all cells contained within a radial bin of width $\Delta r = 0.3 \,\mathrm{kpc}$ and restricted to a cone with semi-opening angle $30^\circ$.
The legend of Fig.~\ref{fig: mix scalars} (right panel) shows the radial average of the ratio between $\dot{M}_\mathrm{IC,cold}$ and the mass outflow rate of the cold phase, $\dot{M}_\mathrm{cold}$. In the case of minimal mixing, as in the fiducial run, one finds $\dot{M}_\mathrm{IC,cold} \simeq \dot{M}_\mathrm{cold}$. In other words, the present-day ($t \simeq 30 \,\mathrm{Myr}$) outflow rate of the cold phase coincides with the outflow rate of gas that was initially cold. This is consistent with the outflowing cold mass being composed of gas that was initially cold in the disc and subsequently accelerated into an outflow, while experiencing negligible mixing with the hot phase, in agreement with $\mathcal{P}_\mathrm{cold,1} \lesssim 1$.
In contrast, when mixing is enhanced, $\dot{M}_\mathrm{IC,cold}$ shows little variation compared to $\dot{M}_\mathrm{cold}$, which instead increases significantly with $f_\mathrm{mix}$. As a result, the average ratio $\langle \dot{M}_\mathrm{IC,cold} / \dot{M}_\mathrm{cold} \rangle$ decreases to $\sim 0.5$ as $f_\mathrm{mix} \to 1$. 
This indicates that although the total amount of initially cold disc gas accelerated into an outflow (regardless of whether it later resides in the hot or cold phase) changes little, the mass outflow rate of the cold phase grows to be about twice as large as that of the \emph{initially} cold phase. Such behavior signals strong mixing between the outward-accelerated disc cold gas and the hot outflow, with a net mass flux from hot to cold that amplifies the cold component of the outflow. This interpretation is consistent with the positive sign of $\dot{M}_\mathrm{net}$ (Figure~\ref{fig: Mnet}) and with $\mathcal{P}_\mathrm{cold,1}$ being $\gtrsim \zeta$, confirming that the cold phase has undergone substantial mixing.

\section{M82 simulations: discussion} \label{sec: discussion}

\subsection{On the degeneracy between mixing and drag in launching cold outflows} \label{sec: degeneracy}

In Figure~\ref{fig: Schneider velocity} we showed that the parameters of the fiducial run are not unique, and that alternative values can also reproduce the velocity profiles reported in \citetalias{Schneider20}. In the fiducial simulation, the contribution of mixing to accelerating the cold outflow is negligible (see the legend in Fig.~\ref{fig:  accelerations}, right panel), while in {\labsim{NoDrag\_fmix0.03}} there is no drag. These two simulations therefore represent extreme cases in which only one of the two source-term processes is active and thus entirely responsible for accelerating the cold phase, demonstrating that either mechanism can independently account for the observed behavior. 

However, the dominance of one process over the other in driving the outflow can affect other properties of the outflow, such as the degree of mixing experienced by the cold phase, as traced by the scalar quantity $\mathcal{P}_\mathrm{cold,1}$. More specifically, when the drag force dominates the launching of the cold outflow, the hot outflow ram pressure strips cold ISM clouds from the disc and accelerates them outward. These clouds then continue traveling with little mixing with the hot phase, consistent with the cold mass outflow rate being nearly equal to the outflow rate of \emph{initially} cold gas  (Fig.~\ref{fig: mix scalars}, right panel), and with $\mathcal{P}_\mathrm{cold,1} \lesssim 1$ (Fig.~\ref{fig: mix scalars}, left panel).
When the role of mixing becomes increasingly important, the mass flux from outflowing hot gas onto ISM cold clouds transfers momentum that can eventually surpass the contribution of drag in generating the cold outflow. While embedded in the outflow, these cold clouds continuously exchange material with the hot phase, so that the cold outflow becomes a mixture of initially cold gas and accreted hot material that has cooled onto them. In this case, $\mathcal{P}_\mathrm{cold,1}$ decreases, approaching $\zeta$ as mixing strengthens and its contribution in accelerating the cold outflow grows. In such cases, most of the radial decline of $\mathcal{P}_\mathrm{cold,1}$ occurs within $R \lesssim 1$~kpc, indicating that mixing is strongest in the inner outflow region, where the bulk of cold-phase acceleration also takes place (Fig.~\ref{fig: accelerations}, right panel). Furthermore, if the mass fluxes between the hot and cold phases are not perfectly balanced but favor a net growth of the cold phase, the cold mass outflow rate can exceed the mass outflow rate of the initially cold gas (Fig.~\ref{fig: mix scalars}, right panel).

However, despite these clear trends, a given value of $0<\mathcal{P}_\mathrm{cold,1} <1$ does not necessarily indicate a dominant process in accelerating the cold outflow. For example, the {\labsim{fmix0.1}} and {\labsim{NoDrag\_fmix0.03}} runs have similar $\mathcal{P}_\mathrm{cold,1}$ profiles and comparable velocity curves\footnote{Compare the {\labsim{fid}} profile with that of {\labsim{fmix0.1}} in Fig.~\ref{fig: mix_vm}, as the {\labsim{NoDrag\_fmix0.03}} profile is nearly identical to that of {\labsim{fid}} (Fig.~\ref{fig: Schneider velocity}).}, yet in the former both drag and mixing contribute almost equally to accelerating the cold outflow, whereas in the latter only mixing drives the acceleration.

More conclusive statements can be made only in extreme regimes. If the composition of the outflowing cold clouds (e.g., their metallicity) matches that of the ISM cold clouds and differs from that of the hot outflow, they must have been accelerated solely by drag, since any contribution from mixing with the hot outflow would have altered their initial composition; this scenario corresponds to $\mathcal{P}_\mathrm{cold,1} \simeq 1$. 
Conversely, if the composition of the cold phase is nearly identical to that of the hot outflow and distinct from the cold ISM, it implies that small seed clouds were accelerated out of the ISM by mixing while acquiring most of their final mass from the hot phase. This corresponds to an \emph{in situ} formation scenario in which $\dot{M}_\textrm{growth} - \dot{M}_\textrm{loss} \lesssim \dot{M}_\textrm{growth}$, unlike in our simulations where growth and loss are nearly balanced, yielding a net hot-to-cold mass flux much smaller than either term (Fig.~\ref{fig: Mnet}, right legend). Such an \emph{in situ} regime is consistent with $\mathcal{P}_\mathrm{cold,1} \simeq 0$. Any intermediate composition of the cold outflow, lying between that of ISM clouds and the hot phase and corresponding to a finite $\mathcal{P}_\mathrm{cold,1}$, would then point to a mixed regime in which both drag and mixing could in principle contribute to launching the cold outflow.

\subsection{Thoughts on the value of the mixing strength $f_\textrm{mix}$}

The comparison between our {\labsim{fid}} and {\labsim{NoDrag\_fmix0.03}} runs with \citetalias{Schneider20} suggests a mixing-strength coefficient in the range $f_\textrm{mix} \sim 0.003$–$0.03$, i.e. 2–3 orders of magnitude smaller than the value estimated by \cite{Fielding22} and \cite{Abruzzo22} from cloud-crushing simulations. In their framework, $f_\textrm{mix}$ quantifies the effective interaction area between cloud and wind, parameterized as $A_\textrm{mix} = f_\textrm{mix},4\pi r_\textrm{cl}^2 \chi^{1/2}$ \citep{Fielding22, Abruzzo22}, with $f_\textrm{mix}=2$ reproducing the simulated mixing rate. 
The lower value of $f_\textrm{mix}$ required in our model to reproduce the high-resolution \citetalias{Schneider20} simulations suggests that additional factors are at play in their setup, causing clouds to interact with the hot wind differently from the behaviour prescribed by our subgrid model, which is based on single-cloud crushing simulations. Since \citetalias{Schneider20} neglect processes such as magnetic fields, cosmic rays and thermal conduction, and we adopt the same cooling function, the origin of this discrepancy must lie elsewhere.  

A plausible explanation is that our approach computes source terms as a simple superposition of single-cloud contributions, while recent studies show that cloud populations exhibit collective behaviors not captured by this additive picture. In multi-cloud systems, compact clouds configurations hinder the formation of extended cloud tails where mixing is most efficient, thereby suppressing mixing relative to more porous distributions \citep{BandaBarragan20}. In addition, hydrodynamic shielding, where upstream clouds protect downstream ones and the large number of clouds obstructs the hot flow, damps the instabilities at cloud–intercloud boundaries that normally drive mixing \citep{BandaBarragan20, Villares24, Seidl25}. As a result, cloud ensembles can survive collectively even when isolated clouds would be rapidly destroyed, owing to this quenched mixing. These collective effects are likely operating in the high-resolution simulations of \citetalias{Schneider20}, while in our clouds population subgrid model they are effectively hidden into a lower value of $f_\textrm{mix}$.

However, we caution that the cold outflow is ultimately driven by the global (cell) source terms, which depend on the product of the mixing strength $f_\textrm{mix}$ and the coefficients $r^{-1}_\textrm{*source*}$ (Eqs.~\ref{Eq: M cold pop loss}, \ref{Eq: M cold pop growth}, \ref{Eq: K pop}). Simplifications and assumptions embedded in the $r^{-1}_\textrm{*source*}$ factors, relative to the cloud populations actually resolved in the \citetalias{Schneider20} simulations, may therefore influence the calibrated value of $f_\textrm{mix}$ required to reproduce their outflow properties.

\subsection{A comparison with \cite{Nguyen24} and \cite{Smith24}} \label{sec: Nguyen}
The decrease in hot-phase velocity with increasing cold-phase mass loading has been previously studied and reported by \cite{Nguyen24} and \cite{Smith24}.

\cite{Nguyen24} systematically investigated the effect of increasing mass transfer from the cold to the hot phase, corresponding to our $\dot{M}_\textrm{loss}$ term, using a simplified form of the source terms considered here. In their approach, the cold phase is not explicitly treated, so there is no $\dot{M}_\textrm{growth}$ term from hot to cold. $\dot{M}_\textrm{loss}$ is constant in time and prescribed as a function of radius. Moreover, the accreted material is injected with zero velocity and negligible thermal energy, and no drag force is included. These differences lead to non-trivial changes in the outflow evolution, but several points of comparison can still be made.
Since the injected mass in their model carries zero momentum, the associated source-term deceleration is more efficient than in our \emph{mixing} term: in the second term of Eq.~(\ref{Eq: acceleration hot}) they effectively have $v_\textrm{hot}$ instead of $v_\textrm{rel}$. However, they lack the drag–induced deceleration term, and their pressure–gradient acceleration increases with larger $\dot{M}_\textrm{loss}$, making it difficult to assess which model predicts a stronger slowdown of the hot wind. Furthermore, because the injected gas has zero velocity, the thermalised kinetic energy from mixing (the second term in Eq.~\ref{Eq: udot pop mix}) is larger, leading to a net heating of the hot phase that grows with $\dot{M}_\textrm{loss}$. This raises both the temperature and the pressure–gradient acceleration of the hot wind. By contrast, in our simulations, stronger mixing rapidly reduces $v_\textrm{rel}$, decreasing both drag and mixing–induced heating, so the hot phase is heated less at higher mixing.
\cite{Nguyen24} also found that increasing $\dot{M}_\textrm{loss}$ can make the hot outflow locally subsonic (Mach number $\mathcal{M}_\textrm{hot} < 1$), triggering a cooling instability that produces cold filaments in the hot wind. In our case, the opposite trend in source-term heating causes $\mathcal{M}_\textrm{hot}$ to increase with stronger mixing. As shown in the left panel of Fig.~\ref{fig: mix Tumix}, the Mach number never falls below unity in any of our simulations, and we do not observe the onset of such an instability.

\cite{Smith24} studied the hot- and cold-phase properties of outflows with varying cold-phase mass loading, i.e. by launching different amounts of cold gas at a fixed velocity at the base of the outflow, without changing any of the source-term parameters (for a comparison between our model and theirs, see Appendix~\ref{app: literature}). An increased amount of cold mass, however, enhances both mixing and drag between the phases, see Eqs. (\ref{Eq: Normalization phi}, \ref{Eq: M cold pop loss}, \ref{Eq: M cold pop growth} and \ref{Eq: K pop}).  
To interpret the differences between our results and theirs, we consider our Eqs.~(\ref{Eq: acceleration hot} and \ref{Eq: acceleration cold}) for the source-term accelerations, which are general and independent of the specific treatment of mass loss, growth, and drag, and can therefore be used to analyze the outcome of both models. 
In our model, increasing $f_\textrm{mix}$ enhances the source–term mixing deceleration of the hot phase, and in the model of \cite{Smith24}, a higher cold–mass loading boosts both the mixing and drag decelerations. In both cases then, the hot–phase velocity decreases (see their Fig.~7).  The two models, however, diverge in their predictions for the cold phase. In our simulations, enhanced mixing also increases the acceleration of the cold phase, leading to faster cold outflows. In contrast, \cite{Smith24} find that stronger cold mass loading reduces the cold-phase velocity. This difference can be explained as follows: when the source-term acceleration of the cold phase is considered (Eq. \ref{Eq: acceleration cold}), increasing mass loading raises both the numerators and denominators by the same factor, in principle leaving the source-term acceleration nearly unchanged. However, stronger mass loading lowers the hot-phase velocity, thereby reducing $v_\mathrm{rel}$; since both acceleration terms scale with $v_\mathrm{rel}$, the net result is a further suppression of the cold-phase acceleration.

\section{Conclusion} \label{sec: conclusion}

In this paper, we have introduced a new model for simulating multiphase galactic outflows using efficient, comparably low-resolution simulations. The framework builds on the two-fluid branch of the \textsc{Arepo} code developed by \citet{Weinberger23}. In our implementation, the cold phase ($T=10^4$ K) is treated as pressureless, and the interaction between the hot and cold phases within each cell is modeled (see Section~\ref{Sec: Model} and Appendix~\ref{app: model}):
i) through source terms that capture the effects of drag and mixing of cold clouds in a hot wind, calibrated against cloud-crushing simulations by \citet{Fielding22}; and
ii) by assuming that clouds within each cell follow a power-law mass distribution $\propto m_\textrm{cloud}^{-2}$, consistent with findings from recent high-resolution galactic wind simulations \citep{Tan24, Warren25}. A schematic overview of the model is shown in Figure~\ref{Fig: scheme}.

We then applied this model to simulations of the multiphase galactic outflow in the starburst galaxy M82. First, we calibrated the model and setup parameters to reproduce the velocity, density, and mass outflow rates of the hot and cold phases as found in the high-resolution simulations of \citetalias{Schneider20}. Building on this fiducial setup, we carried out a suite of simulations in which we varied key model parameters, as summarized in Table~\ref{tab: sim parameters}. Our main results are as follows:

\begin{itemize}

    \item Our model successfully reproduces the multiphase galactic outflow properties of M82 (see Section~\ref{sec: schneider} and Figs.~\ref{fig: Schneider velocity}, \ref{fig: Schneider density}, \ref{fig: Schneider Mdots}). What makes this agreement particularly significant is that, unlike in the \citetalias{Schneider20} simulation, the cold clouds that constitute the cold phase in our runs are neither spatially resolved nor explicitly resolved in their interaction with the hot phase. Instead, both are captured through subgrid modeling as described in Section~\ref{Sec: Model}. This enables us to lower the resolution by orders of magnitude while still obtaining quantitatively consistent predictions for both the resolved hot phase and the unresolved cold phase. Furthermore, as demonstrated in our resolution test (Appendix~\ref{sec: resolution}; see Fig.~\ref{fig: hr_vel}), the model remains robust across a wide range of resolutions, with the subgrid prescriptions retaining their accuracy.

    \item In our simulations, the cold gas is not directly injected together with the hot gas from SNe feedback. Instead, the emergent cold outflow arises solely from the interaction between the hot, stellar-driven wind and the cold phase residing in the galactic disc. We find that the unresolved source-term processes—namely mixing between cold clouds and hot gas, and drag forces (ram pressure exerted by the hot wind on the cold clouds)—contribute comparably to accelerating the cold gas out of the disc and generating a cold outflow. In particular, we show that a run including only drag and our fiducial run (in which both processes are present, but the contribution of mixing to the acceleration of the cold phase is negligible; Fig.~\ref{fig: accelerations}) yield almost identical velocity profiles for both phases (Fig.~\ref{fig: Schneider velocity}). This demonstrates that mixing and drag are degenerate in their ability to drive cold outflows: different relative contributions from the two processes can lead to indistinguishable outcomes when considering velocity profiles. In addition, attempting to disentangle their roles purely from the amount of mixing experienced by the cold phase can be both non-trivial and potentially misleading (see Section~\ref{sec: degeneracy}).

    \item To reproduce the \citetalias{Schneider20} results, our model requires a much smaller mixing-strength coefficient ($f_\textrm{mix} \sim 0.003$–$0.03$) than predicted by cloud-crushing simulations ($f_\textrm{mix} \sim 2$, e.g. \cite{Abruzzo22}, \cite{Fielding22}), likely because collective effects in multicloud systems (e.g. compactness, hydrodynamic shielding) suppress mixing compared to the simple superposition of isolated clouds interacting individually with the hot wind.

    \item Increasing the strength of mixing between phases significantly affects the properties of the multiphase outflow (see Section~\ref{sec: fmix} and Appnedix~\ref{app: fmix}). Stronger mixing enhances the net transfer of mass from the hot to the cold phase, causing the total mass outflow rate to become progressively dominated by the cold component (Fig.~\ref{fig: mix_vm}, bottom row). The hot phase velocity decreases with increasing mixing (Fig.~\ref{fig: mix_vm}, top row) for two reasons (Fig.~\ref{fig: accelerations}): (i) as more mass is transferred from the slower cold phase, the hot phase is progressively decelerated, and (ii) the reduced velocity difference between phases lowers the energy dissipated by ram pressure, which weakens heating of the hot gas and thus the pressure gradient driving its acceleration. The cold phase, on the other hand, experiences progressively stronger acceleration with increasing mixing (Fig.~\ref{fig: accelerations}), to the point where mixing becomes the dominant acceleration channel compared to drag. As a result, the cold gas velocity approaches that of the hot phase more quickly, while the latter is simultaneously damped. Consequently, the cold phase does not necessarily reach higher absolute velocities with stronger mixing, but the velocity contrast between the two phases is substantially reduced.

    \item Different levels of mixing affect not only the kinematics and thermodynamics of the multiphase outflow but also the origin of the outflowing cold component (see Sections~\ref{sec: origin} and \ref{sec: degeneracy}). When mixing is weak, the cold outflow is primarily driven by ram pressure exerted by the hot wind on pre-existing ISM clouds. In this regime, the cold outflowing material originates almost entirely from gas that was already cold in the disc (Fig.~\ref{fig: mix scalars}) and therefore retains the same properties, such as metallicity. By contrast, stronger mixing increases the net mass transfer from the hot to the cold phase, so that a growing fraction of the cold outflowing gas is formed from hot SNe-driven wind that mixes, cools, and accretes onto the cold clouds. In this case, the cold phase acquires a different composition than the disc ISM, enriched by SNe ejecta from the hot wind.

\end{itemize}

Our model relies on several assumptions that introduce intrinsic limitations and may affect the accuracy of our results. The main caveats are:

\begin{itemize}
    \item The single-cloud source term prescriptions, presented in \citep{Fielding22} and based on idealized cloud-crushing simulations \citep[e.g.,][]{Gronke18,Gronke20, Fielding20a, Kanjilal21, Abruzzo22}, assume a spherical cloud impacted by a supersonic hot wind, without magnetic fields, cosmic rays, turbulence in the hot medium, cloud–cloud interactions, or collective effects such as shielding. As such, the predictive power of our model is inherently limited by these idealizations. Nevertheless, our subgrid framework is flexible: updated prescriptions from future, more realistic simulations can be incorporated by modifying the single-cloud source terms.

    \item Beyond these intrinsic limitations, our treatment of a population of clouds introduces further simplifications. We assume a stationary cloud mass function, without explicitly modeling processes such as mergers or fragmentation. Instead, these are implicitly assumed to operate in the background to maintain the stationarity of the mass function.  

    \item The source terms from all clouds in a cell are integrated into a single cell-wide source term, which is then applied to the bulk cell properties. This yields one cold-phase velocity per cell, assigned to all clouds. In reality, each cloud experiences a source term depending on its size and mass, so smaller clouds would accelerate faster than larger ones, leading to velocity divergence within the cell. Our approach at present does not capture this effect, however, multi-species extensions of the model are in principle possible. However, the present model still provides a good estimate for the bulk cold-phase velocity, which is dominated by the larger clouds. 
\end{itemize}

In the context of galaxy formation simulations, our model represents a sub-resolution model for unresolved cold clouds that is inspired in its parametrization by highly idealized cloud crushing simulations, verified by comparison to 1d steady-state solutions (see Appendix~\ref{app:1dtest}) and calibrated against high resolution galactic outflow simulations. And while we have shown that the multi-fluid discretization and inherent lower resolution requirements allow a computationally inexpensive study of the large-scale consequences of different microscopic interactions, its applications are more far-reaching: this model also enables us to embed this type of outflow in larger-scale and cosmological simulations, an ability that will be crucial for the interpretation of observed galactic outflows and their role in galaxy evolution.

\begin{acknowledgements}
The authors thank Max Gronke and Hitesh Kishore Das for the useful discussions.
FB and RW acknowledge funding of a Leibniz Junior Research Group (project number J131/2022). 
\end{acknowledgements}

\bibliographystyle{aa}
\bibliography{aanda} 
\appendix
\section{Further details on the cloud models} \label{app: model}
In this Appendix we provide additional details on the sub-grid cloud models introduced in Section~\ref{Sec: Model}. We begin by examining specific properties of the power-law model: in Section~\ref{app: r-1} we analyze the dependence of the coefficients defined in Eqs.~(\ref{Eq: rloss}, \ref{Eq: rgrowth}, \ref{Eq: rdrag}) on the model parameters, while in Section~\ref{app: fractional} we investigate, within the cloud population that constitutes the cold phase, which subset of clouds provides the dominant contribution to the cell source terms. In Section~\ref{app: delta-ppl} we present a quantitative comparison between the power-law and delta models, and finally, in Section~\ref{app: literature}, we compare these models with similar approaches found in the literature.

\subsection{The $r^{-1}_\textrm{*source*}$ coefficients} \label{app: r-1}
\begin{figure}
    \centering
    \includegraphics[width=1.0\linewidth]{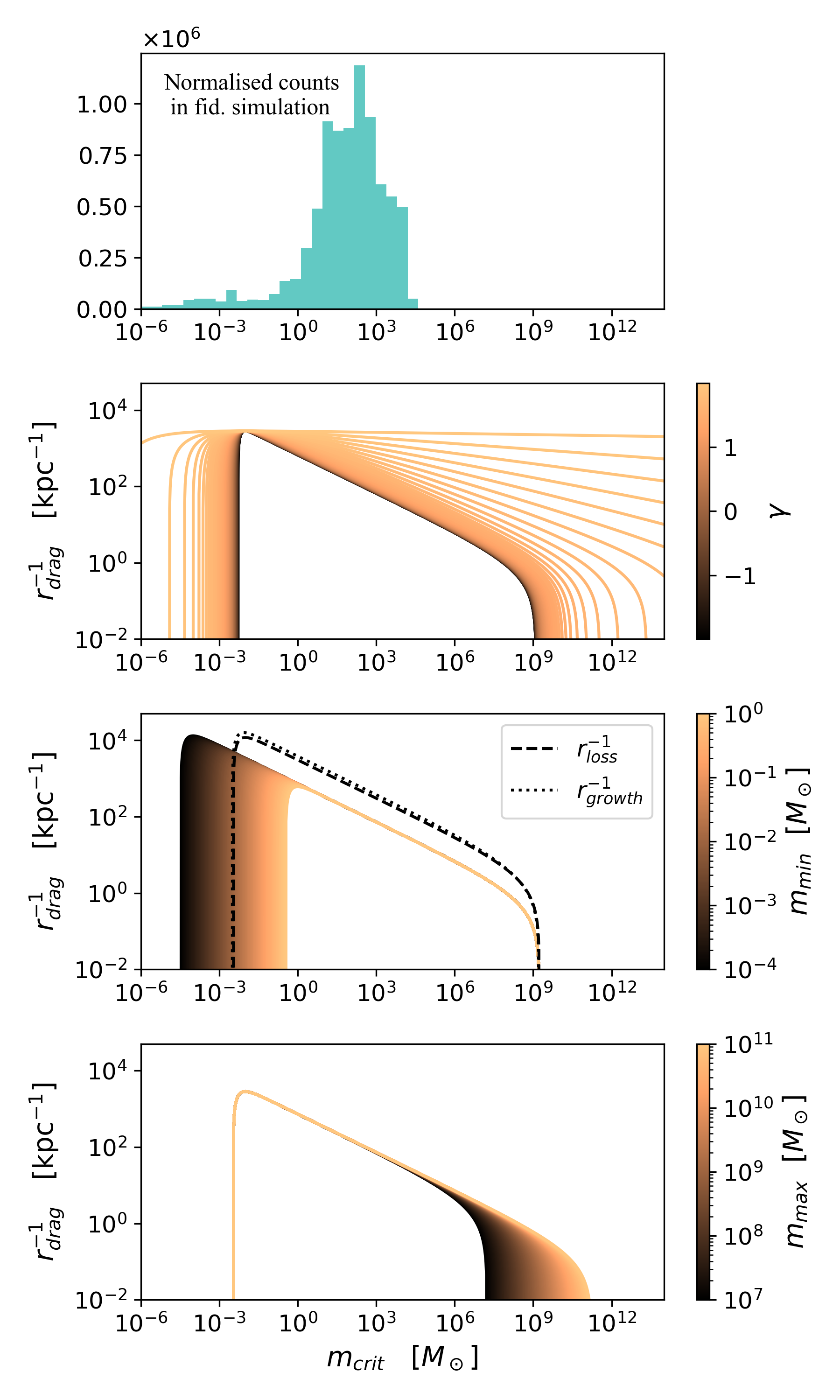}
    \caption{Profiles of the function $r^{-1}_\textrm{drag}(m_\textrm{crit})$ for varying parameter values: $\gamma$ (second panel), $m_\textrm{min}$ (third panel), and $m_\textrm{max}$ (fourth panel). In the second panel, are also shown the profiles of $r^{-1}_\textrm{loss}$ and $r^{-1}_\textrm{growth}$ for the fiducial values ($\gamma = 0$, $m_\textrm{min} = 10^{-2},\mathrm{M}_\odot$, $m_\textrm{max} = 10^9,\mathrm{M}_\odot$). The top panel presents the histogram of $m_\textrm{crit}$ in the outflow of the fiducial simulation over the time interval considered in Section \ref{sec: results}.}
    \label{fig: rdrag}
\end{figure}
The dependence of $r^{-1}_\textrm{drag}$ on $m_\textrm{crit}$ for different choices of model parameters ($\gamma$, $m_\textrm{min}$, $m_\textrm{max}$) is shown in Fig.~\ref{fig: rdrag}. Specifically, the second, third, and fourth panels show $r^{-1}_\textrm{drag}(m_\textrm{crit})$ for varying $\gamma$, $m_\textrm{min}$, and $m_\textrm{max}$, respectively, while keeping the other two parameters fixed at the values used in the fiducial simulation presented in Section~\ref{sec: M82} ($\gamma = 0$, $m_\textrm{min} = 10^{-2}\,\mathrm{M}_\odot$, $m_\textrm{max} = 10^9\,\mathrm{M}_\odot$). The first panel shows a histogram of the $m_\textrm{crit}$ values, which indicates the relevant range of $m_\textrm{crit}$ for the simulations. The third panel also displays the trends of $r^{-1}_\textrm{loss}$ and $r^{-1}_\textrm{growth}$ with $m_\textrm{crit}$. Since their dependence on the model parameters is very similar to that of $r^{-1}_\textrm{drag}$, we omit it in the other panels.  Overall, from Fig.~\ref{fig: rdrag}, we observe that the $r^{-1}$ coefficients, which the source terms are proportional to, are highly sensitive to $m_\textrm{crit}$. These coefficients increase as $m_\textrm{crit}$ decreases, which, according to Eqs.~(\ref{Eq: r_crit}, \ref{Eq: rhocl}), corresponds to conditions with short cooling timescales, low relative velocities between the phases, and low-pressure environments.
The sensitivity to $m_\textrm{min}$ and $m_\textrm{max}$ appears only when $m_\textrm{crit}$ approaches these bounds. In contrast, the dependence on $\gamma$ becomes significant only for $\gamma \gtrsim 1.5$. As $\gamma$ approaches 2, the kink in the mass distribution at $m_\textrm{crit}$ (see Eq.~\ref{Eq: clouds distribution}) gradually disappears, resulting in a flattening of the $r^{-1}$ dependence on $m_\textrm{crit}$.

\subsection{Fractional contributions to the source terms} \label{app: fractional}
\begin{figure*}
    \centering
    \includegraphics[width=1.05\linewidth]{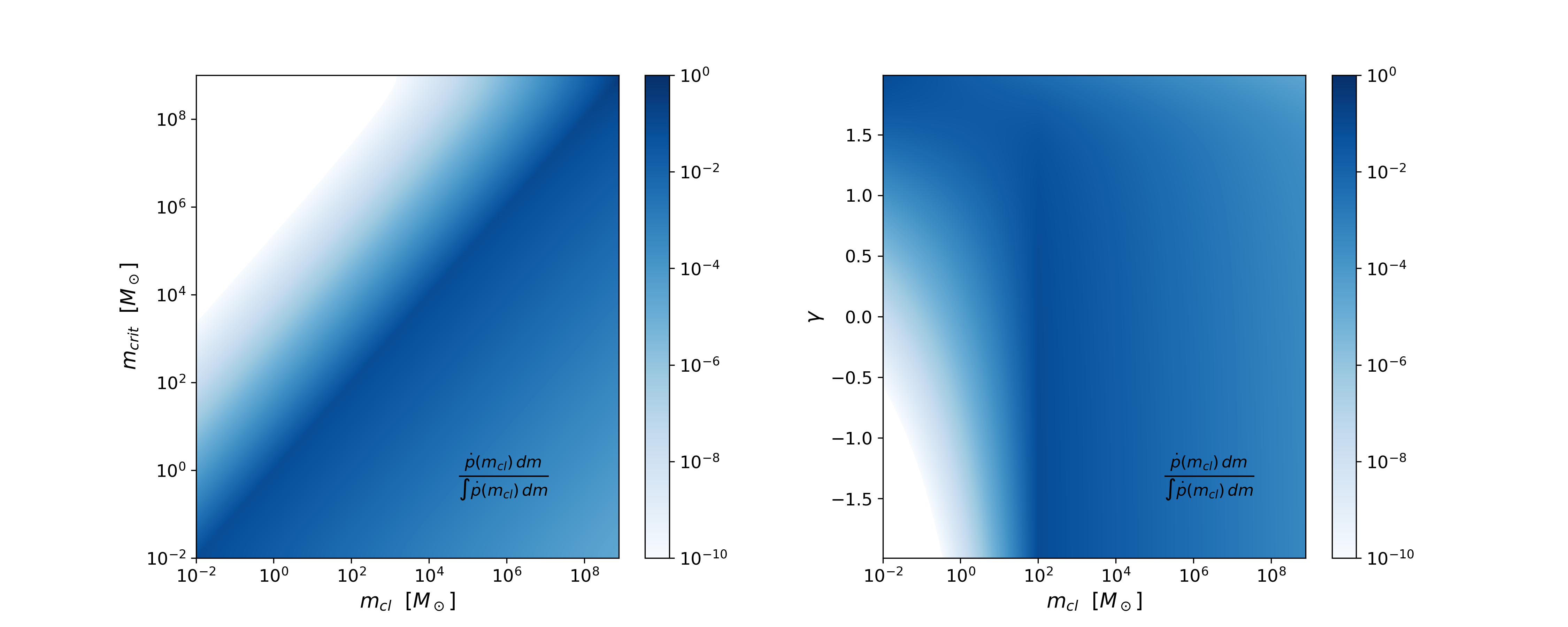}
    \caption{Fractional contribution of clouds in the mass range $(m_\textrm{cl}, m_\textrm{cl} + dm)$ to the total cell drag source term, shown as a function of $m_\textrm{crit}$ and $m_\textrm{cl}$ with $\gamma=0$ (left panel) and as a function of $\gamma$ and $m_\textrm{cl}$ with $m_\textrm{crit} = 10^2$~M$_\odot$ (right panel).}
    \label{fig:Fractional}
\end{figure*}
Here we examine in more detail a specific aspect of the power-law distribution model for clouds, namely: given a population of clouds with masses in the range $(m_\textrm{min}, m_\textrm{max})$, what is the contribution to the total source terms from clouds within the mass interval $(m_\textrm{cl}, m_\textrm{cl} + dm)$? 
Figure~\ref{fig:Fractional} (left panel) shows, through the color bar, the fractional contribution of clouds with mass $m_\textrm{cl}$ to the total drag force, as a function of $m_\textrm{cl}$ ($x$-axis) and of the critical mass $m_\textrm{crit}$ of the distribution ($y$-axis). For this figure, we have adopted fiducial parameters $(\gamma = 0,\,m_\textrm{min} = 10^{-2} \mathrm{M}_\odot,\, m_\textrm{max} = 10^9 \mathrm{M}_\odot)$. The figure demonstrates that the total drag force within a cell is dominated by clouds with mass $m_\textrm{cl} \sim m_\textrm{crit}$, and that the contribution from lower-mass clouds decreases more rapidly than that from higher-mass clouds. This same behavior is observed for the other source terms as well.
These findings are consistent with the results discussed in Section~(\ref{app: r-1}), where we noted that $m_\textrm{crit}$ largely determines the strength of the source terms (Figure~\ref{fig: rdrag}). 
In the right panel of Figure~\ref{fig:Fractional}, we fix $m_\textrm{crit} = 10^2 \mathrm{M}_\odot$ and investigate how the fractional contribution of clouds with mass $m_\textrm{cl}$ varies with the slope $\gamma$. The results indicate that the trend observed in the left panel --- that clouds with $m_\textrm{cl} \sim m_\textrm{crit}$ dominate the source terms --- holds as long as $\gamma \lesssim 1.5$, that is, as long as the distribution retains a clear kink. As $\gamma \rightarrow 2$, this kink disappears, the dependence of the source terms on $m_\textrm{crit}$ vanishes (see also Figure~\ref{fig: rdrag}, second panel), and most of the contribution to the total source terms progressively shifts towards the low-mass end of the distribution.

\subsection{Comparing the delta and power law cloud models} \label{app: delta-ppl}
\begin{figure*}
    \hspace{-0.12\linewidth}
    \includegraphics[width=1.2\linewidth]{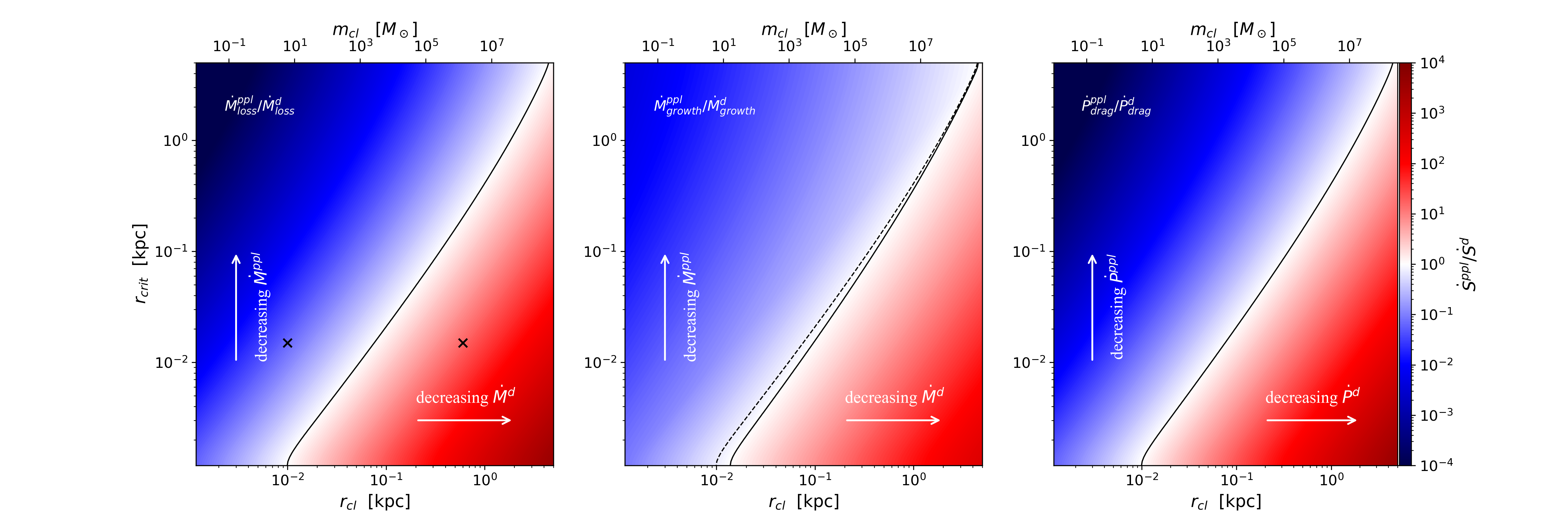}
    \caption{Ratios of the total cell source terms computed using the power-law and delta models for mass loss (first panel), mass growth (second panel), and drag force (third panel). In each panel, the ratio is plotted as a function of the critical radius in the power-law model ($y$-axis) and the cloud size $r_\textrm{cl}$ (bottom $x$-axis) or cloud mass $m_\textrm{cl}$ (top $x$-axis) in the delta model. The solid black lines mark the ratio equal to 1. In the second panel, the dashed black line indicates the location of the equality lines from the first and third panels. The two crosses in the first panel mark the values of $r_\textrm{cl}$ from simulations {\labsim{rcl0.01}} and {\labsim{rcl0.6}}, positioned at approximately the median $r_\textrm{crit}$ of the {\labsim{fid}} run (see Fig.~\ref{fig: others}). }
    \label{fig: comparison models}
\end{figure*}
We now perform a quantitative comparison between the delta distribution model and the piecewise power-law  distribution model. Figure~\ref{fig: comparison models} shows the ratios between the source terms --- mass loss rate $\dot{M}_\textrm{loss}$, mass growth rate $\dot{M}_\textrm{growth}$, and drag momentum $\dot{\mathbf{P}}_\textrm{drag}$ --- for the cloud population within the computational cells, as computed with the two different models. These ratios are plotted as a function of the characteristic radii of the models, namely $r_\textrm{cl}$ for the delta model and $r_\textrm{crit}$ for the power-law model.
For each source term, we observe a consistent trend. At fixed $r_\textrm{crit}$ in the power-law model, increasing $r_\textrm{cl}$ in the delta model results in increasing the ratio $\dot{S}^\textrm{ppl}/\dot{S}^\textrm{d}$. This behavior is expected, as discussed previously (see Section~\ref{sec: delta}), since a population of fewer, larger clouds leads to weaker source terms in the delta model. Similarly, if we fix $r_\textrm{cl}$ and increase $r_\textrm{crit}$, the power-law model becomes dominated by fewer, larger clouds (see Appendix~\ref{app: fractional}), which likewise yields weaker source terms in the power law model, and hence a smaller ratio $\dot{S}^\textrm{ppl}/\dot{S}^\textrm{d}$. In each plot, the solid black lines indicate the pairs $(r_\textrm{cl}, r_\textrm{crit})$ that produce identical source terms. In the second plot, corresponding to $\dot{M}_\textrm{growth}$, we also show (with dashed lines) the equality relation for the other two source terms, which is the same. This equality line differs from that for $\dot{M}_\textrm{growth}$, highlighting that the power-law model cannot be trivially replaced by a delta model with a suitable $r_\textrm{cl}$, as the effective $r_\textrm{cl}$ would vary depending on which source term is being considered.

\subsection{A comparison with other models in the literature} \label{app: literature}

Here, to help situate our new model within the existing literature, we highlight its key similarities and differences compared to two other approaches that also aim to track the evolution of cold gas in a live, sub-grid fashion. The first, and most similar, is the model introduced by \cite{Butsky24}, implemented in the grid-based code \textsc{Enzo}. In their implementation, a two-fluid hydrodynamics framework is employed, where the second (cold) component is treated as a pressureless fluid—that is, pressure forces from the cold gas are neglected in the hydro-solver. This is the same approach adopted in our work. However, it is worth noting that the method presented in \cite{Weinberger23} more generally allows the cold phase to be modeled as a fully compressible fluid \citep[see e.g.][]{Das2024}. Regarding the sub-grid composition of the cold phase, in both \cite{Butsky24} model and our delta model, each resolution element contains equal-size clouds. In our case, the cloud size is set by a free parameter, $r_\textrm{cl}$, while in their case, it is determined by the local thermal and cooling properties of the hot gas within each cell. As in our delta model, the number of clouds per cell is given by $N_\textrm{cl} = M_\textrm{cold}/m_\textrm{cl}$. Moreover, similar to our approach, the hot and cold phases co-evolve in each cell based on the source terms reported by \cite{Fielding22}, which we also employ. Additionally, since the main focus of \cite{Butsky24} is to study cold gas in the circum-galactic medium, they include extra source terms associated with thermal instability, which allow clouds to condense out of the hot phase in the absence of relative velocities.

The second model we compare with is the \textsc{arkenstone} model implemented in \textsc{Arepo} by \cite{Smith24}. In this approach, the cold phase is represented by cold particles that are spawned according to the star formation rate predicted by an ISM model, and are launched out of the ISM with an empirically chosen velocity. These cold particles evolve purely under gravity and interact with the background hot gas through the \cite{Fielding22} source terms, until they are recoupled to the hot phase when their cold mass or the density contrast with the ambient hot gas drop below a specified threshold. Thus, cold-hot source terms are active only as long as these particles remain decoupled from the hot medium.
Their cloud model lies somewhere between our delta and power-law approaches. As in our delta model, each cold particle is assumed to consist of equal-mass clouds, with their mass set by the total cold mass of the particle, $M_\textrm{cold}$ (determined by the spawning prescription), and by the number of clouds per particle, $N_\textrm{cold}$. $N_\textrm{cold}$ is drawn stochastically at the time of spawning, such that the ensemble of clouds \emph{across cold particles} follows a mass distribution $dN/dm \propto m^{-2}$. In general, this method results in different cold particles being characterized by clouds with different masses and sizes, as in \cite{Butsky24}, while in our delta model all cells have the same cloud size, but not necessarily the same cloud mass. 
After spawning, the value of $N_\textrm{cold}$ is then held fixed for the rest of the particle evolution.
In contrast, in our power-law model, the cloud mass distribution is assumed to follow this specific shape within each resolution element at all times. 
Each method has its advantages and limitations. On the one hand, the \textsc{arkenstone} model enforces a power-law cloud mass function at spawning, consistent with the distribution observed by \cite{Tan24}, but this distribution can evolve and deviate from the theoretical expectation as cold particles move through and interact with the hot outflow, due to the differential cloud growth for different cloud masses.
On the other hand, their choice of a single cloud mass per cold particle avoids the problem of diverging velocities among clouds within the same particle. In our power-law model, by contrast, the stationarity of the power-law distribution is built into the subgrid model at all times and locations. As discussed in Section~\ref{sec: ppl}, this implicitly assumes the existence of ongoing fragmentation and merger processes and it is consistend with the stationairty of the distribution observed in \cite{Tan24} and \cite{Warren25}. However, because clouds of different masses coexist within each resolution element, the mass dependence of the single-cloud source terms would cause them to experience different forces and thus diverge in velocity. In practice, we compute the net force on the cold phase in each cell by summing the contributions from all single-cloud terms, which determines the evolution of the bulk cold-phase velocity $\mathbf{v}_\textrm{cold}= \mathbf{P}_\textrm{cold} / M_\textrm{cold}$. We then assume that all clouds move with this bulk cold-phase velocity. This limitation can be addressed in future developments by representing clouds of different sizes by different pressureless fluids, allowing for relative velocities and solving the cloud shattering and coagulation dynamics via mass exchange terms between the pressureless fluids.

\section{1D tests}
\label{app:1dtest}
\begin{figure}
    \includegraphics[width=1\linewidth]{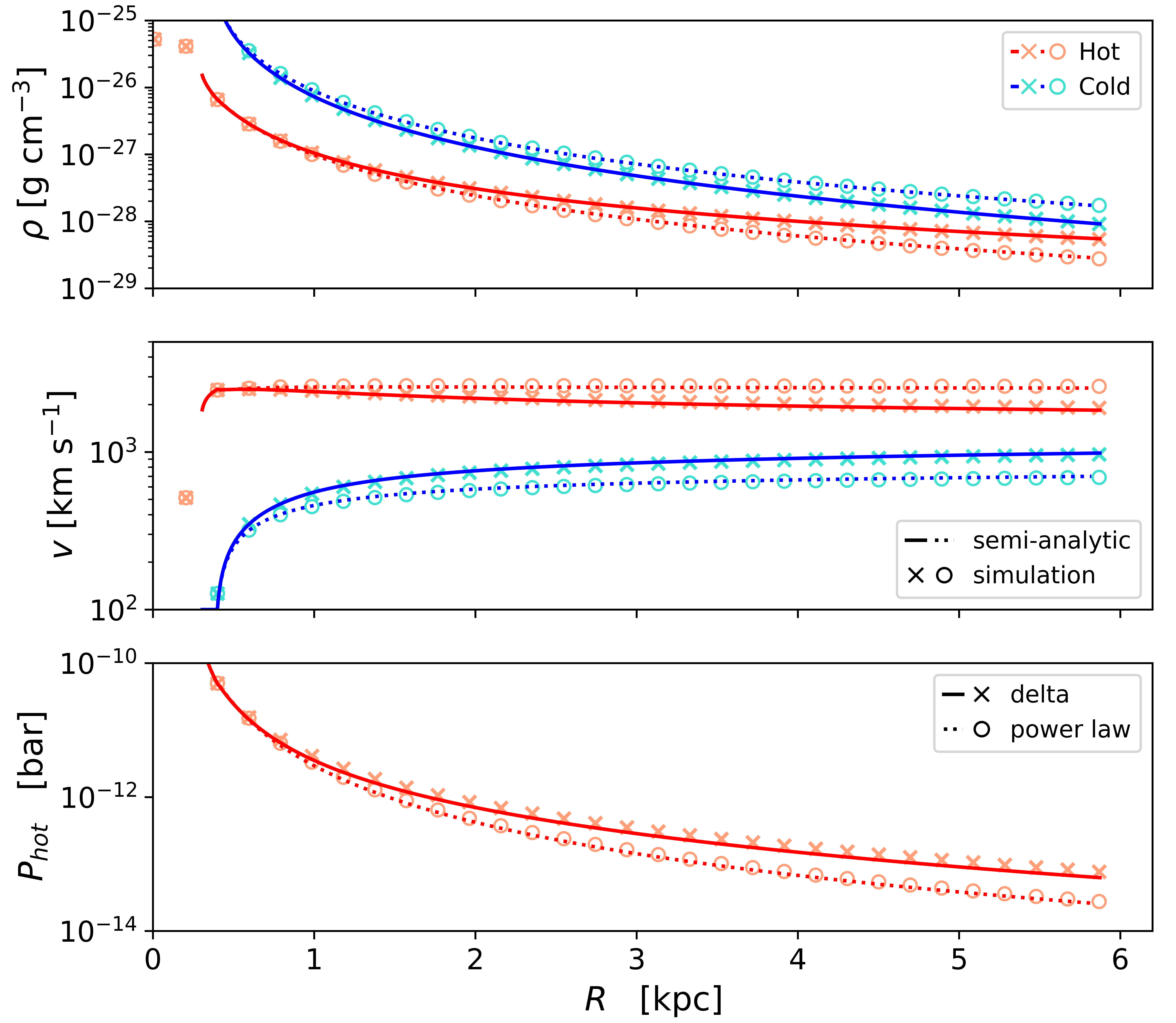}
    \caption{Radial profiles of the asymptotic one-dimensional structure of a multiphase outflow. From top to bottom, the panels show the densities, velocities, and pressures. Curves represent the semi-analytic solutions: solid lines for the delta model and dotted lines for the power-law model. Scatter points represent the simulation results: crosses for the delta model and circles for the power-law model. Blue and turquoise indicate the cold phase, while red and salmon indicate the hot phase.
}
    \label{fig: 1d}
\end{figure}
In this Appendix, we present analytical tests to validate our implementation of the models described in Section~\ref{Sec: Model}. To this end, we consider a one-dimensional (spherically symmetric) stationary problem, for which semi-analytic solutions of the multiphase outflow profiles are available. Specifically, we adopt a setup similar to that discussed in \cite{Fielding22}, characterized by an inner injection of mass and energy in the hot phase within a radius $r_\textrm{inj} = 300$~pc, at rates $\dot{M}_\textrm{inj, hot} = 0.1\,\dot{M}_\star$ and $\dot{E}_\textrm{inj, hot} = \left(10^{51}~\textrm{erg}/100~\mathrm{M}_\odot\right)\cdot\dot{M}_\star$. In addition, we include a cold-phase injection at $r_\textrm{inj, cold} = 400$~pc, with a mass injection rate $\dot{M}_\textrm{inj, cold} = 1.5\,\dot{M}_\textrm{inj, hot}$ and an injection velocity $v_\textrm{inj, cold} = 100$~km~s$^{-1}$. Here we set $\dot{M}_\star = 5~\mathrm{M}_\odot\,\textrm{yr}^{-1}$. 
No gravitational effects are included, and radiative cooling is modeled using a truncated power-law cooling function
\begin{equation}
    \Lambda(T) = \Lambda_0\,T^{-2/3},
\end{equation}
applied only for $T \geq 10^4$~K, where $\Lambda_0 = 1.1\times 10^{-18}~\textrm{erg}~\textrm{K}^{2/3}~\textrm{s}^{-1}~\textrm{cm}^3$.
For these tests, we employed a static one-dimensional mesh with a resolution of 310 cells. Each cell was initialized with a density of $10^{-28}$~g~cm$^{-3}$ and a pressure of $10^{-13}$~bar in the hot phase, and with a density of $10^{-30}$~g~cm$^{-3}$ in the cold phase, both initially at rest. These initial conditions become irrelevant once each cell is reached by the outflow. No refinement or derefinement was enabled. The cold phase was constrained to be in pressure equilibrium with the hot phase, i.e., $\varepsilon_P = 1$, throughout the simulations.  
For the parameters of the power-law models, we adopted those of our fiducial simulation (Table~\ref{tab: sim parameters}), while for the delta model simulation we used $r_\textrm{cl} = 0.1$~kpc. Both simulations were evolved until the asymptotic regime was reached within the simulation domain.   To validate the results against semi-analytic predictions, we employed the publicly available code presented in \cite{Fielding22}, which solves the fluid equations for a stationary, one-dimensional multiphase outflow, including source terms for mixing and drag. To enable a direct comparison with our delta and power-law cloud models, we modified their code to implement our model prescriptions.
More specifically, in their code, \cite{Fielding22} integrate radially the outflow properties $(\rho_\textrm{hot}, v_\textrm{hot}, P_\textrm{hot}, m_\textrm{cl}, v_\textrm{cold})$, where single-cloud source terms are used to evolve $m_\textrm{cl}$ and $v_\textrm{cold}$. These source terms are then incorporated into the hydrodynamic equations of the hot phase to track its evolution. For example, in the density equation, a source term of the form $-n_\textrm{cl}\,\dot{m}_\textrm{cl,mix}$ appears.  
Here, $n_\textrm{cl}$ is indirectly determined from the stationarity condition, being defined as  $n_\textrm{cl} = \dot{N}_\textrm{cl}/4\pi r^2 v_\textrm{cold}$, where $\dot{N}_\textrm{cl}$ is the cloud number flux. This quantity is assumed constant under stationarity and is ultimately set by the initial (at $r_\textrm{inj, cold}$) cloud velocity, mass, and cold-phase mass injection rate. By contrast, in our simulations no stationarity is assumed. Instead, for each cell and at every timestep, the total masses $M_\textrm{cold}$ and $M_\textrm{hot}$ are evolved according to Eq.~(\ref{Eq: umd Mdot}) or (\ref{Eq: ppl Mdot}). To compare with the \cite{Fielding22} code, we modified it so that, instead of tracking the evolution of $m_\textrm{cl}$, we directly track the evolution of\footnote{Note that this is different from the subgrid cloud density $\rho_\textrm{cl}$ of Eq.~(\ref{Eq: rhocl}); here it is defined as the total cold mass per unit volume, equivalent to $M_\textrm{cold}/V_\textrm{cell}$ in the simulations.} $\rho_\textrm{cold}$, using the source terms of Eq.~(\ref{Eq: umd Mdot}) or (\ref{Eq: ppl Mdot}) expressed per unit volume.
Figure~\ref{fig: 1d} compares our test simulations (scatter points) with the semi-analytic models (lines) for both the power-law model (dotted lines and circles) and the delta model (solid lines and crosses). From top to bottom, the panels show the densities, velocities, and pressure of the hot phase. The results demonstrate that, in the asymptotic regime, our simulations accurately reproduce the predicted semi-analytic behavior.
In addition, since in the power-law simulation the typical $r_\textrm{crit} \sim 0.1$--$1$~kpc is smaller than $r_\textrm{cl}$, as discussed in Section~\ref{app: delta-ppl}, the source terms in the power-law model are expected to be weaker. Consistently, we observe more efficient cold-phase acceleration in the delta simulation.

\section{Refinement and derefinement criteria and mesh movement prescription} \label{app: refinement}

To maintain high resolution within the disc—particularly in the inner kiloparsec where supernova mass and energy are injected—while allowing lower resolution in the outflow, we adopt a refinement and derefinement scheme based on the following criterion. Each cell is associated with an elliptical radial coordinate $r_\textrm{ell} =\sqrt{x^2 + y^2 + z^2 (a/c)^2}$ so that all cells within the ellipse $r_\textrm{ell} \leq a$ are assigned a volume of $V_\textrm{disc} = 10^{-5}\,\mathrm{kpc}^3$, while cells outside the ellipse $r_\textrm{ell} \geq d \cdot a$ are assigned a coarser volume of $V_\textrm{out} = 0.1\,\mathrm{kpc}^3$. Here we adopt $a = 1$ kpc, $c = 0.5$ kpc, and $d = 2$. For cells in the intermediate region ($a < r_\textrm{ell} < d \cdot a$), the characteristic cell volume $V_\textrm{int}$ varies smoothly according to the equation:
\begin{equation}
    \log_{10} V_\textrm{int} = \frac{1}{d-1} \Bigr[  \log_{10}\bigl( V_\textrm{crit}/V_\textrm{disc}\bigr) \frac{r_\textrm{ell}}{a} - \log_{10}\bigl( V_\textrm{crit}/V^d_\textrm{disc}\bigr) \Bigl], 
    \label{Eq: V intermediate}
\end{equation}
ensuring a gradual transition in resolution from the disc to the outflow.

We modified the criterion for determining the velocity of the mesh-generating points to incorporate the properties of the cold phase. This adjustment significantly reduces spurious advection errors that arise when most of the cell mass resides in the cold phase. In such cases, the cold phase dominates the mass budget, but the mesh motion would otherwise be tied only to the velocity of the hot phase. 
To avoid this, when the mass in a cell is dominated by the cold phase, i.e., $M_\textrm{cold}/M_\textrm{hot} \geq q$, the mesh velocity is set to match the velocity of the cold phase $\mathbf{v}_\textrm{mesh} = \mathbf{v}_\textrm{cold}$.
Conversely, when the hot phase dominates, i.e., $M_\textrm{cold}/M_\textrm{hot} \leq q^{-1}$, we set
$\mathbf{v}_\textrm{mesh} = \mathbf{v}_\textrm{hot}$.
For intermediate mass ratios, $q^{-1} < M_\textrm{cold}/M_\textrm{hot} < q$, the mesh velocity is smoothly interpolated between the two phases:
\begin{equation}
\mathbf{v}_\textrm{mesh} = \mathbf{v}_\textrm{cold}\left(\frac{q\cdot(M_\textrm{cold}/M_\textrm{hot}) -1}{q^2-1}\right)
+ \mathbf{v}_\textrm{hot}\left(\frac{q\cdot(M_\textrm{cold}/M_\textrm{hot}) -q^2}{1-q^2}\right).
\label{Eq: vmesh}
\end{equation}
This ensures that $\mathbf{v}_\textrm{mesh}$ varies smoothly with the cold-to-hot mass ratio. Throughout our simulations, we adopt $q = 1.2$.

\section{Resolution Tests} \label{sec: resolution}
\begin{figure*}
    \centering
    \includegraphics[width=1.05\linewidth]{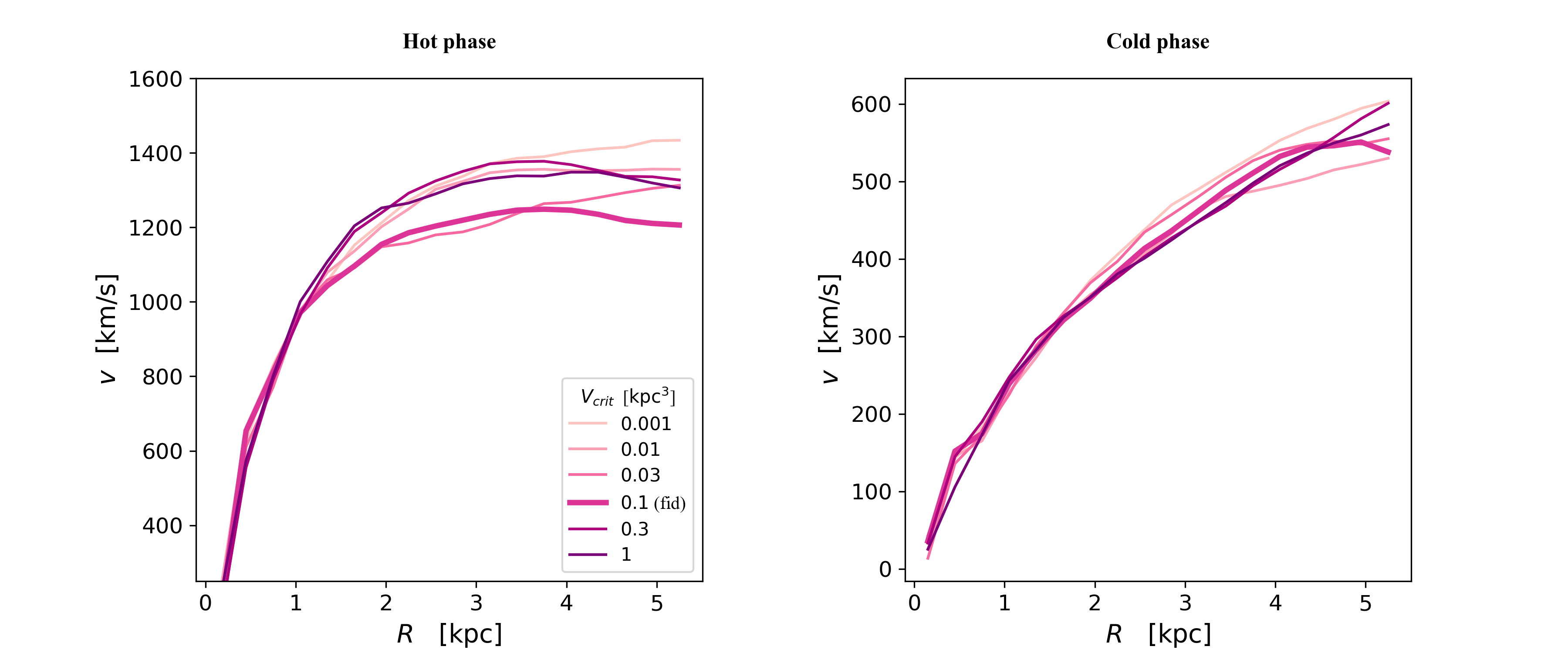}
    \caption{Radial velocity profiles of the hot (left) and cold (right) phases for the fiducial and {\labsim{Vc*}} runs. At each radius, the velocity is computed as the median within radial bins of 0.3~kpc inside a cone with a semi–opening angle of $30^\circ$.}
    \label{fig: hr_vel}
\end{figure*}
In this section, we examine the impact of varying the resolution of the outflow above and below the fiducial value. To this end, we analyse the {\labsim{Vc*}} simulations, which are identical to the fiducial run except for the value of the parameter $V_\textrm{crit}$, which controls the size of the resolution elements in the outflow. The highest–resolution run ({\labsim{Vc0.001}}) resolves the outflow with a factor of 100 more cells than the fiducial simulation, whereas the lowest–resolution run ({\labsim{Vc1}}) uses a factor of 10 fewer cells than {\labsim{fid}}. Overall, the range of resolutions spans three orders of magnitude. In all cases, the gradual increase of cell volumes from their values in the disc $V_\textrm{disc}$ to their maximum value $V_\textrm{crit}$ in the outflow (see Appendix \ref{app: refinement}) ensures that the pressure scale height is always resolved.

Figure~\ref{fig: hr_vel} shows the radial velocity profiles of the hot phase (left) and cold phase (right) for all {\labsim{Vc*}} simulations and the fiducial run. Despite the large differences in resolution, the profiles remain in excellent agreement, with relative deviations from the fiducial case staying within $\lesssim 15\%$. 
This confirms that the numerical method described in Section~\ref{Sec: Model}, and thus the comparison presented in Section~\ref{sec: schneider}, is highly robust to changes in resolution.

\section{Impact of mixing on hot–phase heating and pressure–gradient acceleration} \label{app: fmix}
\begin{figure}
    \centering
    \includegraphics[width=1\linewidth]{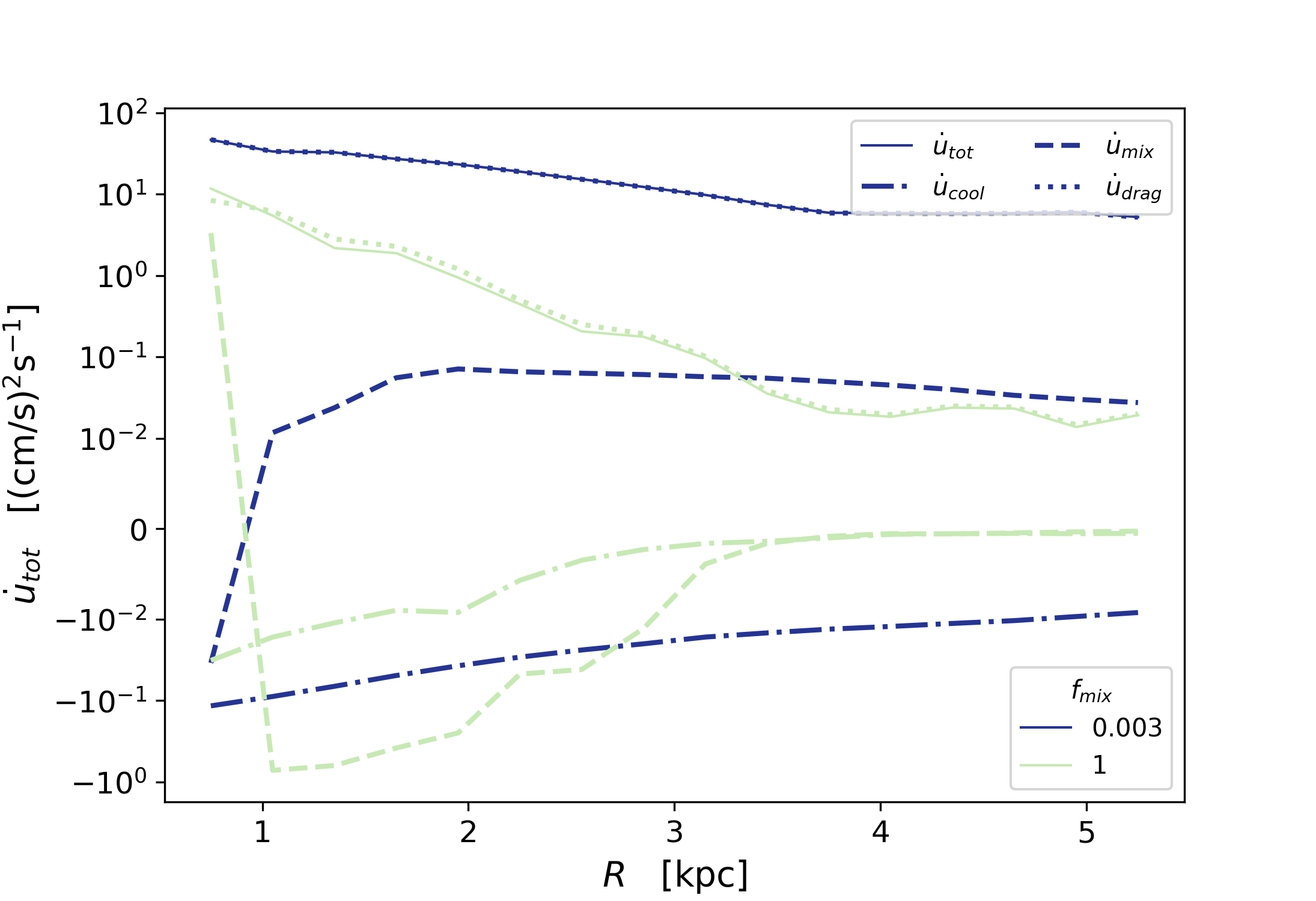}
    \caption{Radial profiles of $\dot{u}_\textrm{tot}$ (solid), $\dot{u}_\textrm{drag}$ (dotted), $\dot{u}_\textrm{mix}$ (dashed) and $\dot{u}_\textrm{cool}$ (dash–dotted) for the {\labsim{fid}} and {\labsim{fmix1}} simulations. }
    \label{fig: utot}
\end{figure}
\begin{figure*}
    \centering
    \includegraphics[width=1.05\linewidth]{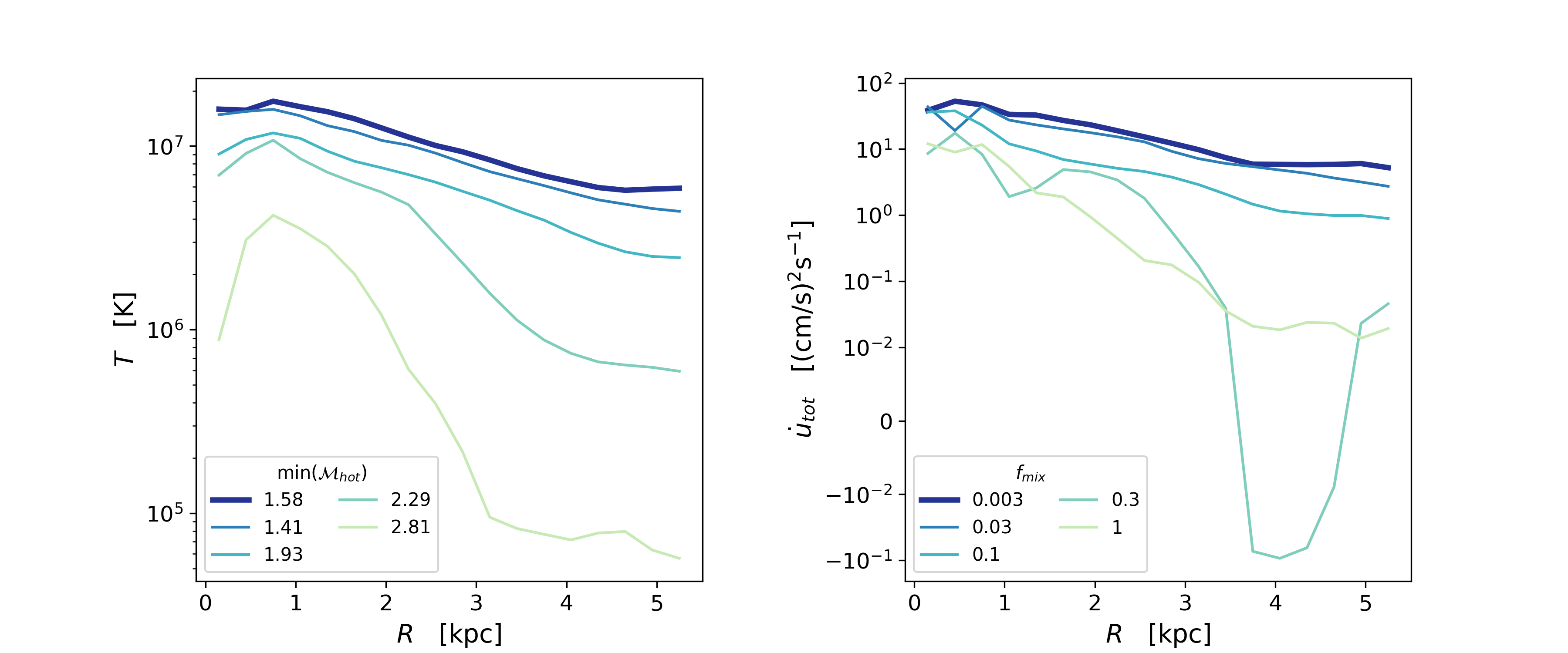}
    \caption{Radial profiles of the hot-phase temperature (left) and of the total hot-phase heating/cooling rate, including both source terms and radiative cooling (right), for the fiducial and {\labsim{fmix*}} simulations. The legend in the left panel also reports the minimum Mach number along the radial profiles for each simulation.}
    \label{fig: mix Tumix}
\end{figure*}
In this Appendix, we explore why increased mixing reduces the pressure–gradient acceleration of the hot wind. Figure~\ref{fig: utot} compares the radial profiles of $\dot{u}_\textrm{tot}$ (Eq.~\ref{Eq: ppl udot}, solid lines) for the simulations with the lowest and highest $f_\textrm{mix}$ ({\labsim{fid}} and {\labsim{fmix1}}), along with the individual contributions: drag heating $\dot{u}_\textrm{drag}$ (Eq.~\ref{Eq: dotu pop drag}, dotted), mixing heating/cooling $\dot{u}_\textrm{mix}$ (Eq.~\ref{Eq: udot pop mix}, dashed), and radiative cooling $\dot{u}_\textrm{cool}$ (dash–dotted).

In both low and high mixing cases, $\dot{u}_\textrm{tot}$ is overwhelmingly dominated by $\dot{u}_\textrm{drag}$, with $\dot{u}_\textrm{mix}$ and $\dot{u}_\textrm{cool}$ providing only minor contributions. For low mixing ($f_\textrm{mix} = 0.003$), $\dot{u}_\textrm{mix}$ is positive because the larger $v_\textrm{rel}$ causes the thermalised relative kinetic energy in Eq.~(\ref{Eq: udot pop mix}) to exceed the (negative) advected thermal energy. At high mixing ($f_\textrm{mix} = 1$), stronger momentum exchange quickly brings the phases to similar velocities, lowering $v_\textrm{rel}$ and making $\dot{u}_\textrm{mix}$ negative, so mixing acts purely as a cooling process for the hot phase.

The right panel of Figure~\ref{fig: mix Tumix} shows $\dot{u}_\textrm{tot}$ for the fiducial run and all {\labsim{fmix*}} simulations. Increasing $f_\textrm{mix}$ consistently reduces $\dot{u}_\textrm{tot}$, mainly because $\dot{u}_\textrm{drag}$ scales as $v_\textrm{rel}^2$ (Eq.~\ref{Eq: dotu pop drag}) and stronger mixing lowers $v_\textrm{rel}$ through enhanced momentum transfer (Fig.~\ref{fig: accelerations}). While $\dot{u}_\textrm{mix}$ also declines and eventually becomes negative with higher mixing, its contribution remains secondary as seen above, except for the {\labsim{fmix0.3}} run where cooling due to mixing is as important as drag heating ($\dot{u}_\textrm{mix} \sim - \dot{u}_\textrm{drag}$). The net effect of increasing mixing is reduced heating of the hot phase, a lower temperature (left panel of Fig.~\ref{fig: mix Tumix}), and thus a weaker pressure–gradient force, which becomes less able to counteract the deceleration from mixing.


\section{Varying $f_\textrm{cold}$ and comparing power law and delta models} \label{sec: other}

\begin{figure*}
    \centering
    \includegraphics[width=1.1\linewidth]{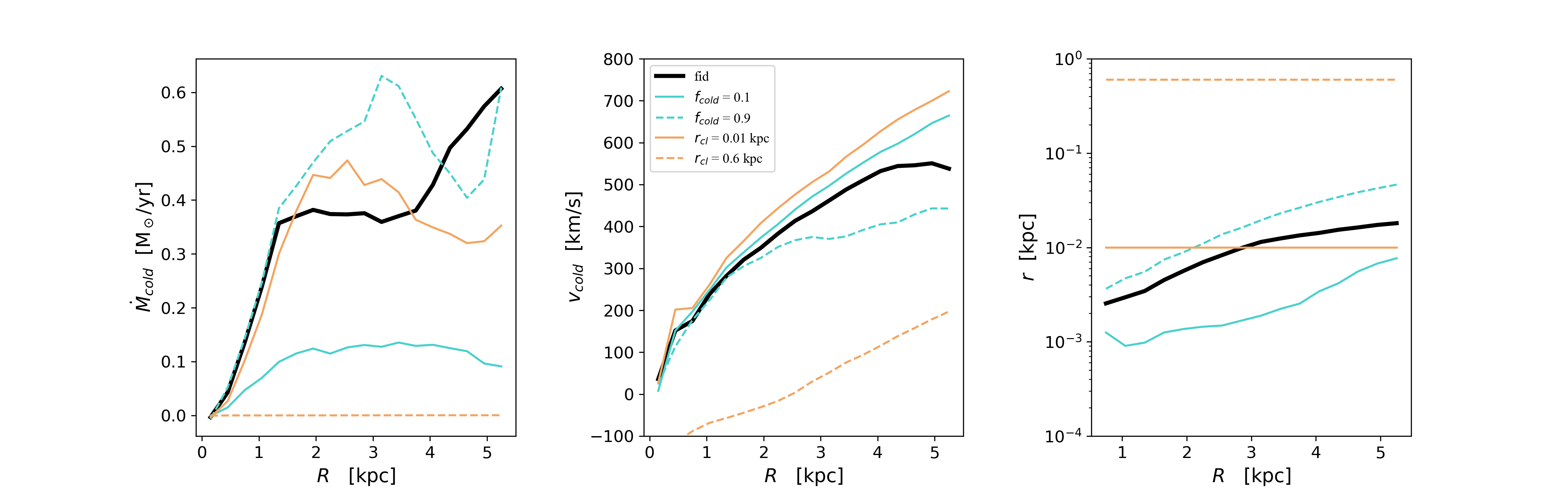}
    \caption{Radial profiles of the cold mass outflow rate (left), cold–phase velocity (center), and characteristic cloud radius (right). For the {\labsim{rcl*}} runs this corresponds to $r_\textrm{cl}$, while for all other runs it corresponds to $r_\textrm{crit}$. Results are shown for the fiducial, {\labsim{rcl*}}, and {\labsim{fcold*}} simulations.
 }
    \label{fig: others}
\end{figure*}

In this section we investigate the impact of varying the initial cold gas fraction $f_\mathrm{cold}$ and of adopting the delta-function cloud model with different cloud radii $r_\mathrm{cl}$, in comparison to our fiducial simulation. For this purpose, we analyze the {\labsim{fcold*}} and {\labsim{rcl*}} runs. Figure \ref{fig: others} shows, for all these simulations, the cold mass outflow rates (left panel), the cold-phase velocity (central panel), and the critical radii $r_\mathrm{crit}$ (or, for the {\labsim{rcl*}} runs, the cloud size $r_\mathrm{cl}$; right panel).

When the cold gas fraction is varied, both the drag force and the mixing-rate source terms change proportionally, since they scale with the cold mass $M_\mathrm{cold}$ in each cell (Eqs.~\ref{Eq: Normalization phi}, \ref{Eq: M cold pop loss}, \ref{Eq: M cold pop growth}, \ref{Eq: K pop}). In particular, a higher cold gas fraction results in a larger cold-phase mass outflow rate, and this larger outflowing mass in turn strengthens the source terms. In the cold-phase acceleration (Eq.~\ref{Eq: acceleration cold}), the cold mass appears in both the denominator and numerator (in the source terms) and thus cancels out. However, stronger source terms also entail stronger source–term heating\footnote{
In the {\labsim{fmix*}} simulations, only the mixing was enhanced; this reduced $v_\textrm{rel}$ and therefore suppressed drag heating, which was otherwise the dominant contribution. In contrast, when $f_\textrm{cold}$ is increased, both the mixing and drag source terms grow because $\mathcal{N} \propto M_\textrm{cold}$. This increase is not compensated by the accompanying reduction in $v_\textrm{rel}$ in our simulations, leading to stronger overall heating as the amount of cold gas rises.}, which lengthens the cooling timescale of the mixing layers and ultimately increases the critical cloud size (Eq.~\ref{Eq: r_crit}), which, according to \ref{app: fractional} is the typical size of clouds that dominate the source terms. A population of fewer, larger clouds yields weaker source terms (Section~\ref{sec: delta}), which in our power law model is encoded through smaller $r_\mathrm{*source*}^{-1}$ coefficients at higher $r_\textrm{crit}$. Since larger clouds correspond to smaller $r_\mathrm{*source*}^{-1}$ values, the net effect is a less efficient acceleration of the cold phase.
Conversely, if the outflowing cold phase is reduced due to a smaller initial amount of cold gas, the weaker source terms result in less source-term heating, leading to smaller $r_\mathrm{crit}$ values and consequently larger $r_\mathrm{*source*}^{-1}$ coefficients for the source terms. This, in turn, enhances the acceleration of the cold phase\footnote{In \cite{Smith24}, the increase in cold mass loading—analogous to increasing $f_\textrm{cold}$—also leads to a reduction of the cold phase velocity. However, while in our case this effect arises from enhanced heating yielding larger cloud sizes, which weaken the source terms responsible for accelerating the cold phase, in their case the dominant factor (as discussed in Section~\ref{sec: Nguyen}) was the strong reduction of the hot phase velocity, not observed in our {\labsim{fcold*}} runs. This reduction dampens $v_\textrm{rel}$ and, consequently, the efficiency of the source terms accelerating the cold phase.}.
Indeed, Figure \ref{fig: others} shows that the {\labsim{fcold0.9}} simulation is characterized by larger cloud critical radii compared to the fiducial run (right panel), and, consistently, the cold phase reaches lower velocities (central panel). Conversely, when the initial cold mass fraction is smaller, the opposite occurs: in the {\labsim{fcold0.1}} run the critical radii are reduced relative to the fiducial case, and the cold phase is accelerated more efficiently.

For the simulations using the delta model ({\labsim{rcl*}} runs), we adopted two different cloud radii $r_\textrm{cl}$ chosen to lie well above and below the typical $r_\textrm{crit}$ of the fiducial simulation, as marked with crosses in Figure \ref{fig: comparison models}. This allows the different outcomes relative to the fiducial run to be interpreted in terms of the expected ratios of the source terms shown in that figure. In the {\labsim{rcl0.01}} run, the cold phase is populated by many small, equal–sized clouds compared to the fiducial case, resulting in stronger source terms. As a consequence, the cold phase is more efficiently accelerated by the source–term (Figure \ref{fig: others}, central panel). By contrast, in {\labsim{rcl0.6}} the cloud population consists of fewer, much larger equal–sized clouds, yielding comparatively weaker source terms, to the extent that the cold outflow fails to be launched (Figure \ref{fig: others}, central and left panels).

\end{document}